\documentclass[aps,prd,onecolumn,superscriptaddress,showpacs,amsmath,amssymb]{revtex4}

\usepackage{graphicx}% Include figure files
\usepackage{dcolumn}% Align table columns on decimal point
\usepackage{setspace}
\usepackage{subfigure}
\usepackage{slashbox}

\usepackage{lineno}

\begin{document}

\newcommand{\mtop}{\mbox{$M_{\rm top}$}}
\newcommand{\ttbar}{\mbox{$t\bar{t}$}}
\newcommand{\thetastar}{\mbox{$\theta^*$}}
\newcommand{\costhetastar}{\mbox{$\cos\thetastar$}}
\newcommand{\Mlb}{\mbox{$M_{lb}$}}
\newcommand{\Et}{\mbox{$E_T$}}
\newcommand{\Pt}{\mbox{$p_T$}}
\newcommand{\Flong}{\mbox{$F_0$}}
\newcommand{\Fplus}{\mbox{$F_+$}}
\newcommand{\met}{\mbox{$\protect \raisebox{0.3ex}{$\not$}\Et$}}

\bibliographystyle{revtex}

\vspace*{1.5cm}

\title{Measurement of the $p \bar p \to t\bar t$ Production Cross Section and the Top Quark Mass  at $\sqrt{s}=1.96$\,TeV in the All-Hadronic Decay Mode.}

\affiliation{Institute of Physics, Academia Sinica, Taipei, Taiwan 11529, Republic of China} 
\affiliation{Argonne National Laboratory, Argonne, Illinois 60439} 
\affiliation{Institut de Fisica d'Altes Energies, Universitat Autonoma de Barcelona, E-08193, Bellaterra (Barcelona), Spain} 
\affiliation{Baylor University, Waco, Texas  76798} 
\affiliation{Istituto Nazionale di Fisica Nucleare, University of Bologna, I-40127 Bologna, Italy} 
\affiliation{Brandeis University, Waltham, Massachusetts 02254} 
\affiliation{University of California, Davis, Davis, California  95616} 
\affiliation{University of California, Los Angeles, Los Angeles, California  90024} 
\affiliation{University of California, San Diego, La Jolla, California  92093} 
\affiliation{University of California, Santa Barbara, Santa Barbara, California 93106} 
\affiliation{Instituto de Fisica de Cantabria, CSIC-University of Cantabria, 39005 Santander, Spain} 
\affiliation{Carnegie Mellon University, Pittsburgh, PA  15213} 
\affiliation{Enrico Fermi Institute, University of Chicago, Chicago, Illinois 60637} 
\affiliation{Comenius University, 842 48 Bratislava, Slovakia; Institute of Experimental Physics, 040 01 Kosice, Slovakia} 
\affiliation{Joint Institute for Nuclear Research, RU-141980 Dubna, Russia} 
\affiliation{Duke University, Durham, North Carolina  27708} 
\affiliation{Fermi National Accelerator Laboratory, Batavia, Illinois 60510} 
\affiliation{University of Florida, Gainesville, Florida  32611} 
\affiliation{Laboratori Nazionali di Frascati, Istituto Nazionale di Fisica Nucleare, I-00044 Frascati, Italy} 
\affiliation{University of Geneva, CH-1211 Geneva 4, Switzerland} 
\affiliation{Glasgow University, Glasgow G12 8QQ, United Kingdom} 
\affiliation{Harvard University, Cambridge, Massachusetts 02138} 
\affiliation{Division of High Energy Physics, Department of Physics, University of Helsinki and Helsinki Institute of Physics, FIN-00014, Helsinki, Finland} 
\affiliation{University of Illinois, Urbana, Illinois 61801} 
\affiliation{The Johns Hopkins University, Baltimore, Maryland 21218} 
\affiliation{Institut f\"{u}r Experimentelle Kernphysik, Universit\"{a}t Karlsruhe, 76128 Karlsruhe, Germany} 
\affiliation{High Energy Accelerator Research Organization (KEK), Tsukuba, Ibaraki 305, Japan} 
\affiliation{Center for High Energy Physics: Kyungpook National University, Taegu 702-701, Korea; Seoul National University, Seoul 151-742, Korea; SungKyunKwan University, Suwon 440-746, Korea} 
\affiliation{Ernest Orlando Lawrence Berkeley National Laboratory, Berkeley, California 94720} 
\affiliation{University of Liverpool, Liverpool L69 7ZE, United Kingdom} 
\affiliation{University College London, London WC1E 6BT, United Kingdom} 
\affiliation{Centro de Investigaciones Energeticas Medioambientales y Tecnologicas, E-28040 Madrid, Spain} 
\affiliation{Massachusetts Institute of Technology, Cambridge, Massachusetts  02139} 
\affiliation{Institute of Particle Physics: McGill University, Montr\'{e}al, Canada H3A~2T8; and University of Toronto, Toronto, Canada M5S~1A7} 
\affiliation{University of Michigan, Ann Arbor, Michigan 48109} 
\affiliation{Michigan State University, East Lansing, Michigan  48824} 
\affiliation{University of New Mexico, Albuquerque, New Mexico 87131} 
\affiliation{Northwestern University, Evanston, Illinois  60208} 
\affiliation{The Ohio State University, Columbus, Ohio  43210} 
\affiliation{Okayama University, Okayama 700-8530, Japan} 
\affiliation{Osaka City University, Osaka 588, Japan} 
\affiliation{University of Oxford, Oxford OX1 3RH, United Kingdom} 
\affiliation{University of Padova, Istituto Nazionale di Fisica Nucleare, Sezione di Padova-Trento, I-35131 Padova, Italy} 
\affiliation{LPNHE, Universite Pierre et Marie Curie/IN2P3-CNRS, UMR7585, Paris, F-75252 France} 
\affiliation{University of Pennsylvania, Philadelphia, Pennsylvania 19104} 
\affiliation{Istituto Nazionale di Fisica Nucleare Pisa, Universities of Pisa, Siena and Scuola Normale Superiore, I-56127 Pisa, Italy} 
\affiliation{University of Pittsburgh, Pittsburgh, Pennsylvania 15260} 
\affiliation{Purdue University, West Lafayette, Indiana 47907} 
\affiliation{University of Rochester, Rochester, New York 14627} 
\affiliation{The Rockefeller University, New York, New York 10021} 
\affiliation{Istituto Nazionale di Fisica Nucleare, Sezione di Roma 1, University of Rome ``La Sapienza," I-00185 Roma, Italy} 
\affiliation{Rutgers University, Piscataway, New Jersey 08855} 
\affiliation{Texas A\&M University, College Station, Texas 77843} 
\affiliation{Istituto Nazionale di Fisica Nucleare, University of Trieste/\ Udine, Italy} 
\affiliation{University of Tsukuba, Tsukuba, Ibaraki 305, Japan} 
\affiliation{Tufts University, Medford, Massachusetts 02155} 
\affiliation{Waseda University, Tokyo 169, Japan} 
\affiliation{Wayne State University, Detroit, Michigan  48201} 
\affiliation{University of Wisconsin, Madison, Wisconsin 53706} 
\affiliation{Yale University, New Haven, Connecticut 06520} 
\author{T.~Aaltonen}
\affiliation{Division of High Energy Physics, Department of Physics, University of Helsinki and Helsinki Institute of Physics, FIN-00014, Helsinki, Finland}
\author{A.~Abulencia}
\affiliation{University of Illinois, Urbana, Illinois 61801}
\author{J.~Adelman}
\affiliation{Enrico Fermi Institute, University of Chicago, Chicago, Illinois 60637}
\author{T.~Affolder}
\affiliation{University of California, Santa Barbara, Santa Barbara, California 93106}
\author{T.~Akimoto}
\affiliation{University of Tsukuba, Tsukuba, Ibaraki 305, Japan}
\author{M.G.~Albrow}
\affiliation{Fermi National Accelerator Laboratory, Batavia, Illinois 60510}
\author{S.~Amerio}
\affiliation{University of Padova, Istituto Nazionale di Fisica Nucleare, Sezione di Padova-Trento, I-35131 Padova, Italy}
\author{D.~Amidei}
\affiliation{University of Michigan, Ann Arbor, Michigan 48109}
\author{A.~Anastassov}
\affiliation{Rutgers University, Piscataway, New Jersey 08855}
\author{K.~Anikeev}
\affiliation{Fermi National Accelerator Laboratory, Batavia, Illinois 60510}
% ...
\collaboration{CDF Collaboration\footnote{With visitors from $^a$University of Athens, 15784 Athens, Greece, 
$^b$University of Bristol, Bristol BS8 1TL, United Kingdom, 
$^c$University Libre de Bruxelles, B-1050 Brussels, Belgium, 
$^d$Cornell University, Ithaca, NY  14853, 
$^e$University of Cyprus, Nicosia CY-1678, Cyprus, 
$^f$University College Dublin, Dublin 4, Ireland, 
$^g$University of Edinburgh, Edinburgh EH9 3JZ, United Kingdom, 
$^h$University of Heidelberg, D-69120 Heidelberg, Germany, 
$^i$Universidad Iberoamericana, Mexico D.F., Mexico, 
$^j$University of Manchester, Manchester M13 9PL, England, 
$^k$Nagasaki Institute of Applied Science, Nagasaki, Japan, 
$^l$University de Oviedo, E-33007 Oviedo, Spain, 
$^m$University of London, Queen Mary College, London, E1 4NS, England, 
$^n$University of California Santa Cruz, Santa Cruz, CA  95064, 
$^o$Texas Tech University, Lubbock, TX  79409, 
$^p$University of California, Irvine, Irvine, CA  92697, 
$^q$IFIC(CSIC-Universitat de Valencia), 46071 Valencia, Spain, 
}}
\noaffiliation

\begin{abstract}
We report the measurements of the $t\bar t$ production cross section and of the top quark mass using 
 1.02 fb$^{-1}$ of  $p\bar p$ data collected with the CDF\, II detector at the Fermilab Tevatron. 
We select events with six or more jets on which a number of kinematical requirements are imposed by means of
 a neural network algorithm. At least one of these jets must be identified as initiated by a $b$-quark candidate by the reconstruction of  a secondary vertex. 
The cross section is measured to be 
$\sigma_{\ttbar}=8.3\pm 1.0(\textnormal{stat.})^{+2.0}_{-1.5}(\textnormal{syst.})\pm 0.5(\textnormal{lumi.})$\,pb,
which is consistent with the standard model prediction. 
The top quark mass of $174.0\pm 2.2(\textnormal{stat.})\pm 4.8(\textnormal{syst.})$ GeV/c$^2$ is 
derived from a likelihood fit incorporating reconstructed mass distributions representative of signal and background.
\end{abstract}

%uncomment for the PRD
\maketitle

\section{\label{sec:Intro}Introduction}

The measurement of the top quark properties allows one to verify the consistency of the standard model.
At the Fermilab Tevatron Collider top quarks are produced mostly in pairs and 
the measurement of the $\ttbar$ cross section tests the next-to-leading-order QCD calculations.
Moreover,  accurate measurements of the top quark mass and of the mass of the $W$ boson 
 provide constraints on the mass of the hypothetical standard model Higgs boson\,\cite{higgs}. 

At the Tevatron center-of-mass energy, $\sqrt{s} = 1.96$\,TeV, the predicted $\ttbar$ production cross 
section is 6.7\,pb\,\cite{topxs} for an assumed  top quark mass of  175~GeV/$c^2$, with a total theoretical uncertainty
 of approximately 15\%,  due to the choice of renormalization and factorization scales.
The top quark decays into a $W$ boson and a $b$ quark almost 100\% of the time; the $W$ boson subsequently 
decays to either a quark-antiquark pair or a lepton-neutrino pair. The resulting final states are then usually distinguished
 by the number of charged energetic leptons ($e$ or $\mu$) and the number of jets.

In this analysis, we examine events characterized by a multijet topology
 and no energetic lepton (``all-hadronic'' mode). This $t\bar t$ final state has the advantage of a large branching ratio ($\approx 4/9$) and of 
having no undetected top decay physical observables in the final state.
In addition, discrepancies in top quark cross section and mass measurements between this and other decay channels
could indicate contributions from physics beyond the standard model\,\cite{kane}.
The major challenge of this channel is the 
very large QCD multijet background which dominates the signal by three orders 
of magnitude after the application of the online trigger selection. 
To improve the signal-to-background ratio (S/B), requirements based on the kinematical and topological 
characteristics of standard model $\ttbar$ events are expressed in terms of a  neural network and applied to the data.
This neural network selection reaches a S/B of about 1/12, improving the S/B by $60 \%$ 
with respect to the selection based on kinematical cuts used in previous analyses\,\cite{tophadPRD, ideogram}.
The neural network selection is followed by the requirement of  jets identified as originating from $b$ 
quarks using a secondary vertex $b$-tagging algorithm, achieving a S/B  of about 1/2.
 
The purity of the selected sample enables the clear observation of $t\bar t$ candidates and the
measurement of the $t\bar t$ production cross section and of the top quark mass.
Top quark events result
in final states with a number of $b$-tagged jets larger than in inclusive QCD multijet production, so we use the
excess of such tags to measure the $t\bar t$ production cross section.
A reconstructed top quark mass is then determined from a kinematical fit of the six leading jets in the event to a $t\bar t$ final state. The distribution of  reconstructed top quark masses is then compared to the distributions expected from background and  $t\bar t$ events simulated at various values of the top quark mass, to obtain the value which best describes the data.

Given the theoretical uncertainties on the production cross section for  
 events generated with N partons at tree level, a more accurate background estimate is obtained from the data themselves 
(``data-driven'') rather than from 
theoretical prediction of cross section and simulations.

The CDF and D\O\, Collaborations previously measured the $\ttbar$ production cross section and the top quark mass 
in the all-hadronic channel\,\cite{tophadPRL} using datasets with integrated luminosities of 
approximately 110\,pb$^{-1}$ collected at $\sqrt{s}=1.8$\,TeV during the Tevatron Run I (1992 - 1996). 
More recent Run II measurements of the cross 
section\,\cite{tophadPRD} and mass\,\cite{ideogram} have been performed by CDF using 311\,pb$^{-1}$ collected  with 
the CDF\, II detector at $\sqrt{s}=1.96$\,TeV (March 2002 - August 2004).
The results reported here are based on the data taken between March 2002 and February 2006, corresponding 
to an integrated luminosity of $1.02$~fb$^{-1}$.
These measurements complement  other recent $t\bar t$ cross section and top quark mass determinations by CDF\,\cite{ljetsSVX,ljetsKIN,metjetPRL,cdfmass} and D\O\,\cite{d0xsec,d0xsec2,d0mass} in other final states.

 The organization of the paper is as follows: Section\,\ref{sec:detector} contains a brief description
 of the CDF\,II detector. The trigger and the neural-network-based sample selection are described in
 Section\,\ref{sec:dataset}. The $b$-tagging algorithm and its efficiency for identifying $b$ jets are
 described in Section\,\ref{sec:btag}. 
In Section\,\ref{sec:bkg} the data-driven method we use for estimating the background from experimental multijet data is applied
 and the related systematic uncertainties are evaluated. Section\,\ref{sec:optim} describes the 
optimization of the kinematical selection and the associated efficiency. 
The $\ttbar$ production cross section
is presented in Section\,\ref{sec:xs}. The reconstruction of the top quark mass and the corresponding background and signal
distributions are described in Section\,\ref{sec:chi2}; the method for fitting
 these distributions is discussed in Section\,\ref{sec:like}. Section
 \ref{sec:masssyst} summarizes the expected contributions to the systematic uncertainty on the top quark mass measurement, while 
Section \ref{sec:mass} describes the top quark mass measurement. Finally, cross section
 and mass measurements are  summarized in Section\,\ref{sec:fine}.

\section{\label{sec:detector}The CDF\,II Detector}

The CDF\,II detector\,\cite{CDFdetector} is an azimuthally and forward-backward symmetric apparatus designed 
to study $p\bar p$ collisions at the Tevatron. A cylindrical coordinate system as described
 in\,\cite{coordinate} is used.
 The detector consists of a magnetic spectrometer surrounded by calorimeters and muon chambers. 
The charged particle tracking system is immersed in a 1.4~T solenoidal magnetic field with axis parallel to the beamline. A set of silicon microstrip detectors provides charged particle tracking in
 the radial range from 1.5 to 28~cm, while 
a 3.1~m long open-cell drift chamber, the central outer tracker (COT), covers the radial range
 from 40 to 137 cm. 
%The COT provides up to 96 measurements of the track position with alternating 
%axial and $\pm$2$^{\circ}$-stereo superlayers, each superlayer having 12 layers of sense wires.
In combination the silicon and COT detectors provide excellent tracking up to about pseudorapidities\,\cite{coordinate} $|\eta|\le 1.1$, with decreasing coverage up to $|\eta| \le 2.0$.
Segmented electromagnetic and hadronic calorimeters 
surround the tracking system and measure the energy of interacting particles. 
The electromagnetic and hadronic calorimeters are lead-scintillator and iron-scintillator sampling devices, respectively, covering the range $|\eta|\le 3.6$. 
They are segmented in the central region ($|\eta|<1.1$) in towers of 15$^\circ$ in azimuth and 0.1 
in $\eta$, and in the forward region ($1.1<|\eta|<3.6$) in towers of 7.5$^\circ$ for
 $|\eta|<2.11$ and 15$^\circ$ for $|\eta|>2.11$.  
The electromagnetic calorimeters\,\cite{ecal,pem} are instrumented with proportional and 
scintillating strip detectors that measure the transverse profile of electromagnetic showers
 at a depth corresponding to the expected shower maximum.
Drift chambers located outside the central hadronic calorimeters and behind a 60~cm iron shield 
detect muons with $|\eta| \le 0.6$\,\cite{CMU}. Additional drift chambers and scintillation  
counters detect muons in the region $0.6<|\eta|<1.5$. Gas Cherenkov counters\,\cite{CLC} with a coverage of $3.7<|\eta|<4.7$ measure the average number of inelastic $p\bar p$ collisions and thereby 
determine the luminosity.
 
\section{\label{sec:dataset}Multijet Event Selection}

The all-hadronic final state of $\ttbar$ events is characterized by the presence of at least 
six  jets from the decay of the two top quarks. 
Collisions are selected in real time using a  multijet trigger, relying on calorimeter information, which was specially developed to collect
 the events used in this analysis. 
Subsequently, jets are identified off-line by grouping clusters of energy in the calorimeter using a  fixed-cone 
algorithm with a  radius 0.4 in  $\eta - \phi$ space\,\cite{jets}. The typical jet transverse energy\,\cite{coordinate} resolution 
is approximately $(0.1\times E_T +1.0)$ GeV\,\cite{jetresol}, where $E_T$ is the jet transverse energy in GeV. After a preliminary selection of multijet events, a neural network selection based on relevant  kinematical variables is used to provide the most precise cross section and mass measurements. 

\subsection{Multijet Trigger}
The CDF trigger system has three levels.
The first two levels consist of 
special-purpose electronics and the third one of conventional digital processors.
For triggering purposes the calorimeter granularity is simplified to a $24\times 24$ grid in
 $\eta - \phi$ space with each ``trigger tower'' spanning approximately 15$^\circ$ in $\phi$ and 0.2
 in $\eta$ covering one or two physical towers.  
At level 1, the jet trigger requires a single tower with transverse energy $E_T^{tow}\ge 10$~GeV.
 At level 2 we require that the total transverse energy, summed over all the trigger towers, 
$\sum E_T^{tow}$, be $\ge 175$~GeV, and the presence of four or more clusters of calorimeter towers, each cluster with
transverse energy $E_T^{cls}\ge~15$~GeV. 
In order to maintain an acceptable trigger rate, as the peak instantaneous luminosity increased, the $\sum E_T^{tow}$ threshold has been increased over the course of the data taking, 175 GeV being the latest value. This threshold has been applied, off-line, to the whole data set.
Finally, the third trigger level confirms the level 2 selection using a more accurate 
determination of the jet energy, requiring four or more reconstructed jets with 
$\Et\ge 10$~GeV. %(with respect to $z=0$). 
A total of  4\,340\,143 events satisfy the trigger requirements, with an efficiency of about 58\% for inclusive $\ttbar$ events, and of 80\% for all-hadronic $\ttbar$ events.  The efficiency has been estimated using Monte Carlo generated $t\bar t$ events, described later.
This corresponds to the  S/B of approximately 1/1100 for the theoretical cross section of $6.7$~pb, where the background is 
represented by the multijet events themselves.

\subsection{Preselection and Topology  Requirements}

Events satisfying  the trigger requirements are reconstructed in terms of their final state observables (tracks, vertices, charged leptons, jets). We retain only those events that are well contained in detector acceptance, requiring the primary event vertex\,\cite{vertex} to lie inside the luminous region ($|z|<60$~cm).
The jet energies are corrected for detector response and multiple interactions. 
First, we take into account the $\eta$-dependence of  detector response 
and energy loss in the uninstrumented regions.
After a small correction for the extra energy deposited by multiple collisions in the same
 beam-beam bunch crossing, a correction for calorimeter non-linearity is applied so that 
the jet energies correspond,  on average, to the energy of all the particles incident on  the jet cone.
All systematic uncertainties for the individual corrections are added in quadrature to obtain
the total uncertainty on the estimate of the initial parton energy. This uncertainty
 goes from 8\% to 3\% with jet transverse energy increasing from 15 GeV to 50 GeV, and remains approximately constant at  
3\% above 50 GeV\,\cite{JESNIM}. 

For this analysis, each jet is required to have $\Et\ge 15$~GeV and $|\eta|\le 2$ after 
all corrections have been applied. 
In order to remove the events from the $\ttbar$ leptonic channels,
 we veto events containing any well identified high-$\Pt$ electrons and muons as defined in\,\cite{ljetsKIN}, and require that 
$\frac{\met}{\sqrt{\sum\Et}}$ be $<3$\,$\sqrt{{\rm GeV}}$\,\cite{metjetPRL}, 
where the missing transverse energy, $\met$\,\cite{met}, is computed with reference to the detector origin and is corrected for
 any identified muons and the position of the $p\bar p$ collision point, while $\sum\Et$ is obtained by summing 
the $E_T$'s of all the selected jets. At this stage, called ``preselection'', simulations show that  the fraction of leptonic events amounts to about
14\% of all accepted $t\bar t$ events.
To avoid overlaps between jets we require jet pairs to be
separated by at least 0.5 units in the $\eta - \phi$ space.
About 3.5 million events pass these preselection requirements (S/B $\sim 1/1000$).
Finally, we define the topology of the signal region by selecting events with a number 
of jets $6\le N_{\rm jets}\le 8$ to optimize the signal fraction. 506\,567 events 
pass this additional requirement with an expected S/B of approximately 1/370, i.e. about 0.3\%; the remaining 
events with lower jet multiplicity have much smaller values of S/B and are used as control regions. The residual fraction of leptonic  $\ttbar$ events amounts to only 5\% of all accepted events.

\subsection{Neural-Network-based Kinematical Selection}

\begin{figure}[htbp]
\centering
\includegraphics[width=8.0cm]{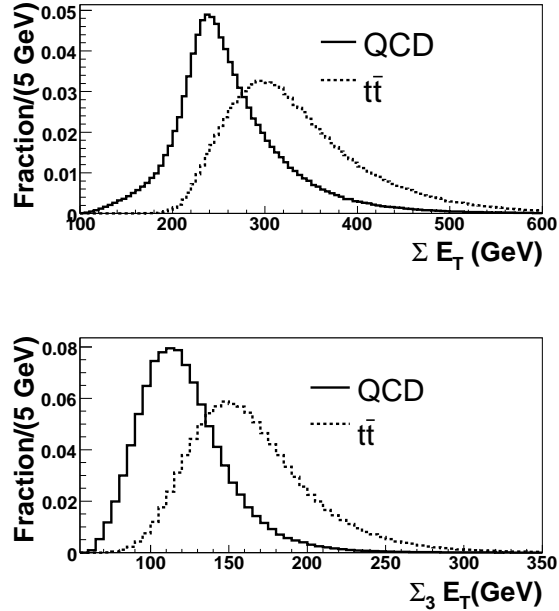}
\caption{$\sum E_T$ (top) and  $\sum_3 E_T$ (bottom) distributions in QCD multijet (solid histogram) and $\ttbar$ Monte Carlo 
(dashed histogram) events with $6\le N_{\rm jets}\le 8$. 
All histograms are normalized to unity.} \label{fig:kinsel1a}
\end{figure}

\begin{figure}[htbp]
\centering
\includegraphics[width=8.0cm]{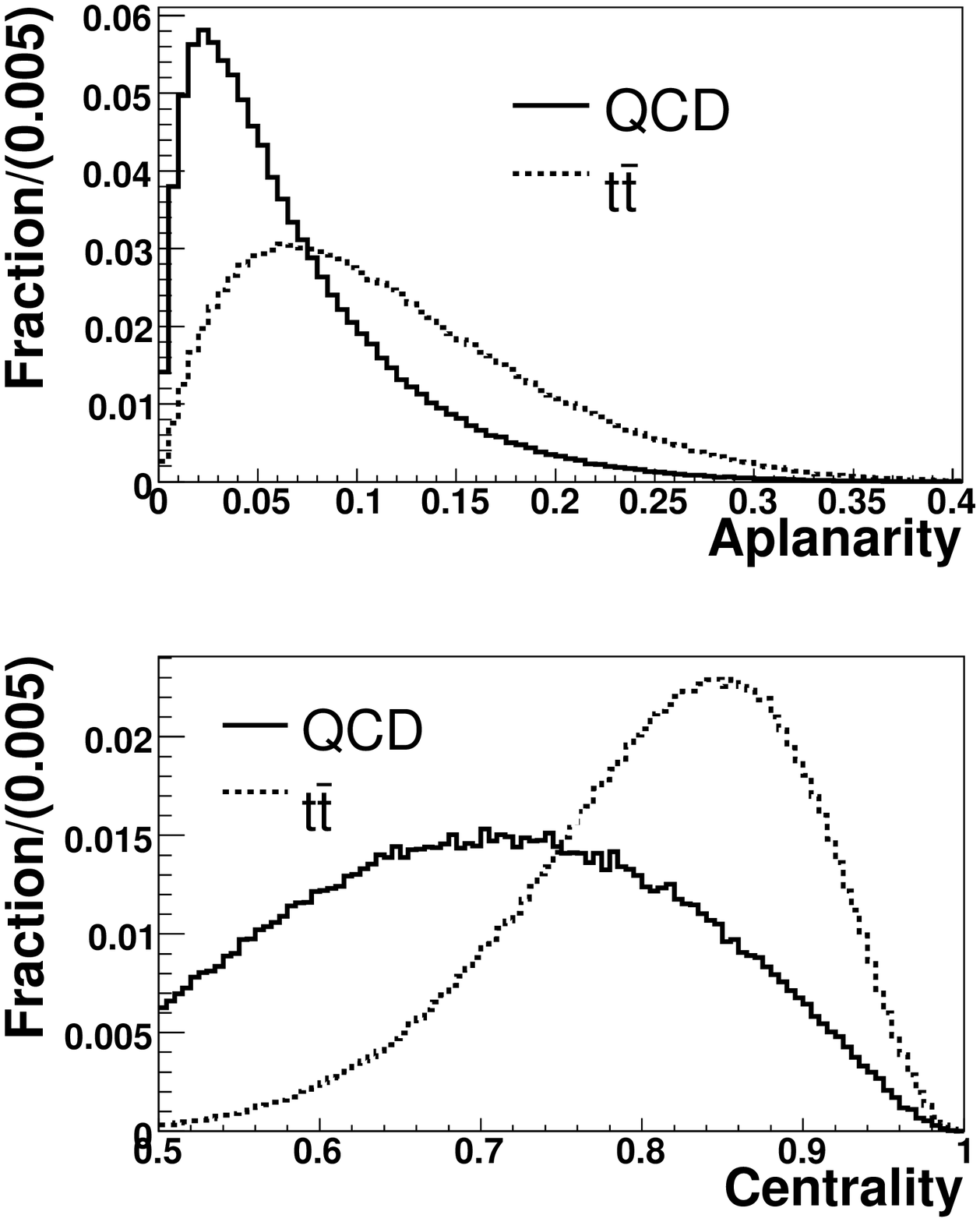}
\caption{Aplanarity (top) and  centrality (bottom) distributions in QCD multijet (solid histogram) and $\ttbar$ Monte Carlo 
(dashed histogram) events with $6\le N_{\rm jets}\le 8$. 
All histograms are normalized to unity.} \label{fig:kinsel1b}
\end{figure}

\begin{figure}[htbp]
\centering
\includegraphics[width=8.0cm]{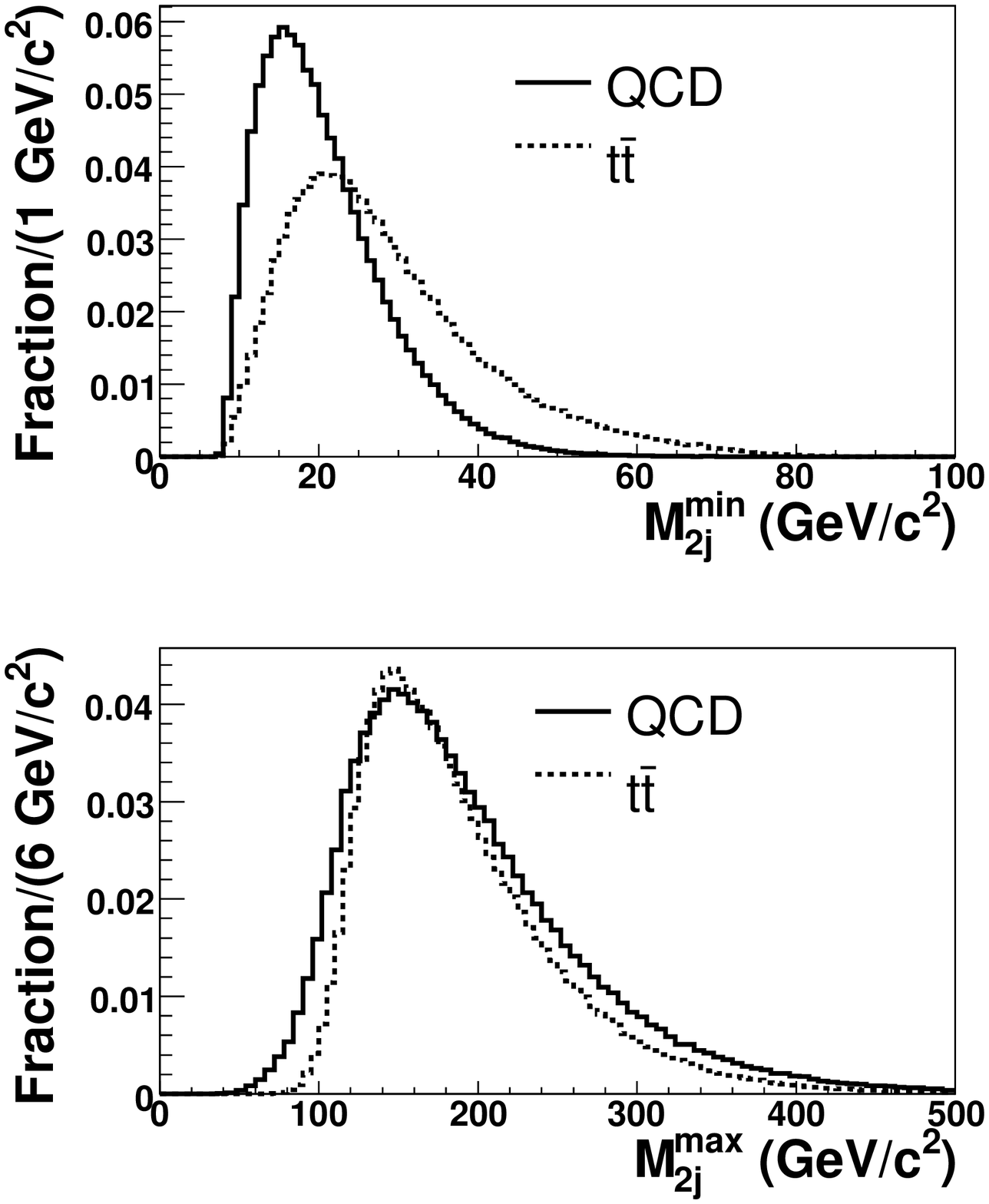}
\caption{$M_{2j}^{min}$ (top) and $M_{2j}^{max}$ (bottom) distributions in QCD multijet  (solid histogram) and $\ttbar$ Monte Carlo 
 (dashed histogram) events with $6\le N_{\rm jets}\le 8$.
  All histograms 
are normalized to unity.} \label{fig:kinsel2a}
\end{figure}

\begin{figure}[htbp]
\centering
\includegraphics[width=8.0cm]{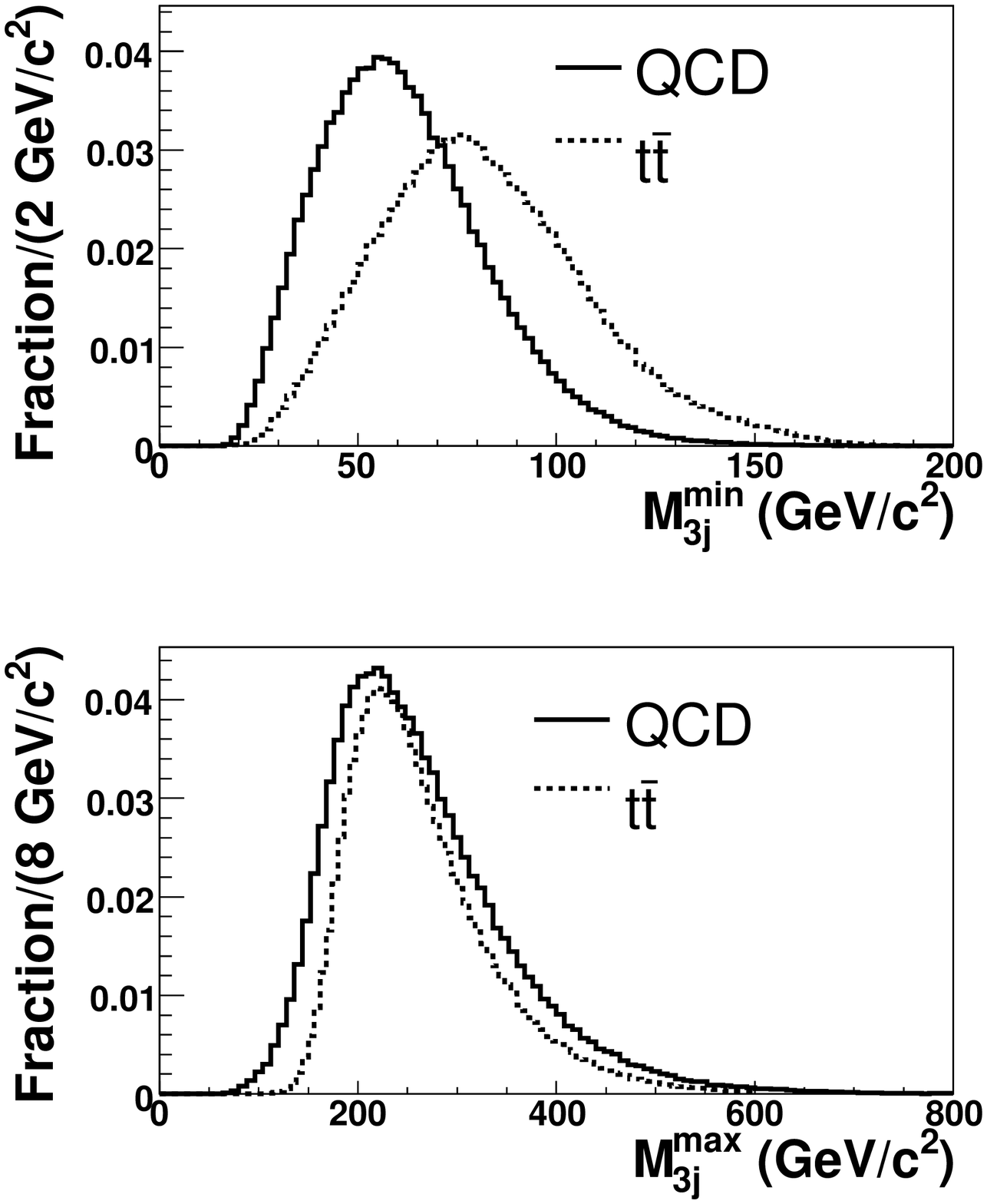}
\caption{$M_{3j}^{min}$ (top) and $M_{3j}^{max}$ (bottom) distributions in QCD multijet  (solid histogram) and $\ttbar$ Monte Carlo 
 (dashed histogram) events with $6\le N_{\rm jets}\le 8$.
  All histograms 
are normalized to unity.} \label{fig:kinsel2b}
\end{figure}

\begin{figure}[htbp]
\centering
\includegraphics[width=8.0cm]{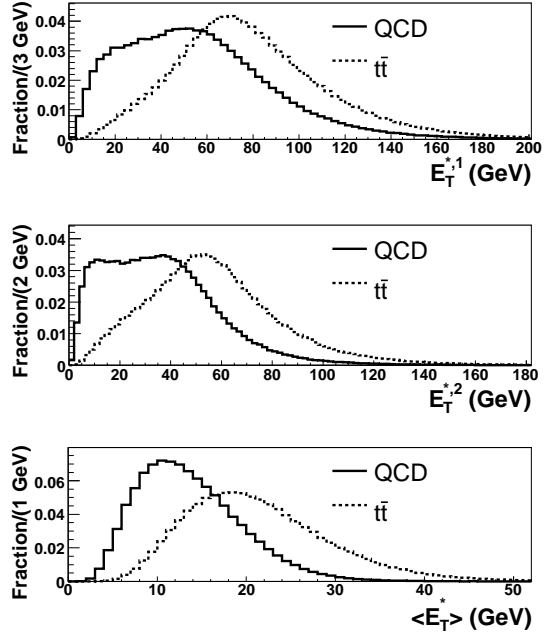}
\caption{Kinematical distributions in QCD  multijet  (solid histogram) and $\ttbar$
 Monte Carlo (dashed histogram) events with $6\le N_{\rm jets}\le 8$. From top: 
 $E_T^{\star, 1}$, $E_T^{\star, 2}$ and $\langle E_T^\star\rangle$. All histograms are
 normalized to unity.} \label{fig:kinsel3}
\end{figure}

\begin{figure}[htbp]
\centering
\includegraphics[width=8.0cm]{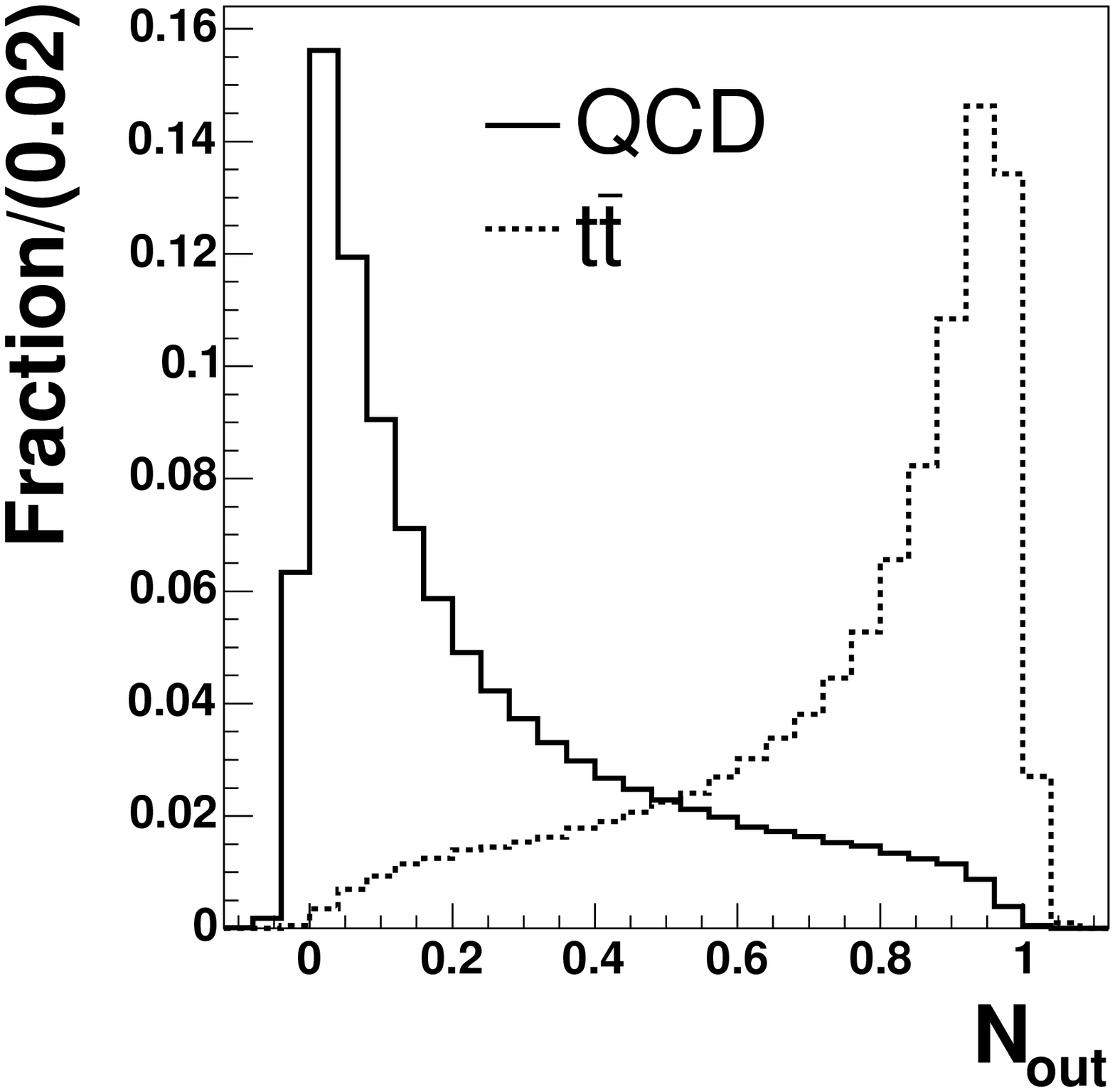}
\caption{Neural network output, $N_{out}$, for QCD multijet (solid histogram) and 
$\ttbar$ Monte Carlo (dashed histogram) events with $6\le N_{jets}\le 8$. Histograms are normalized to unity. The neural network implementation 
that we use in the 
{\em TMultiLayerPerceptron} produces
an output which is not strictly bound between 0 and 1.} \label{fig:nnout}
\end{figure}

In order to further improve the S/B
 we use  a neural
 network approach to recognize in more detail the features of signal and 
background events, including correlations between the kinematical variables which enter as input nodes in the network. 
We thus expect a better separation between signal and background relative to the former technique\,\cite{tophadPRD} where 
correlations were not fully considered. 
The network uses the {\sc mlpfit} package\,\cite{mlp} as implemented by {\sc root}\,\cite{root} through 
the {\em TMultiLayerPerceptron} class. 
We define a kinematical selection based on dynamical and topological properties of the candidate event. 
 The number of variables used should allow the 
best possible description of the event properties, but, at the same time, too many input
 variables can worsen the performance, given the limited training statistics. 
As a guideline we studied different neural network 
configurations, in terms of inputs and hidden nodes, 
adding a few variables at a time, looking for the best performance in terms of largest S/B.
The first quantities considered are those used in\,\cite{tophadPRD}: the total transverse
 energy of the jets, $\sum \Et$; the quantity $\sum _3\Et\equiv \sum \Et - \Et^1 - \Et^2$,
 obtained by removing the contribution of the two jets with the highest $\Et$;
the centrality, defined as $C = \frac{\sum E_T}{\sqrt{\hat{s}}}$, where $\sqrt{\hat{s}}$ is 
the invariant mass of the multijet system; and the aplanarity $A$, defined as 
${A}=\frac{3}{2}{\cal{Q}}_1$, ${\cal{Q}}_1$ being the smallest of the three normalized 
eigenvalues of the sphericity tensor, $M^{ab}= \sum _j P_{j}^a P_{j}^b$, calculated in the 
center-of-mass system of all jets, where the indices $a$ and $b$ refer to the spatial 
components of the jet four-momentum $P_j$.
In addition, we consider the dynamical properties of dijet and trijet systems through the 
use of the minimum and maximum value of the invariant mass among all possible jet 
permutations: $M_{2j}^{min}$, $M_{2j}^{max}$, $M_{3j}^{min}$ and $M_{3j}^{max}$. 
We obtain another set of discriminating variables combining the transverse 
energy of the jets with their emission direction, represented by the angle $\theta^\star$ between the 
jet direction, as measured  in the center-of-mass frame of all
 jets, and the proton beam axis. The variable $\cos\theta^\star$ has been shown\,\cite{costh} to have
discriminating power against the background, and we use here 
the quantity $E_T^{\star}=E_T\sin^2\theta^\star$ which tends to have larger values
 in the signal in comparison to the background events; this effect is enhanced for  the jets  with
higher $E_T$. The variables we choose as additional inputs to the neural network are 
then $E_T^{\star, 1}$ and $E_T^{\star, 2}$ for the two highest-$E_T$ jets, and
 $\langle E_T^\star\rangle$ defined as the geometric mean over the remaining $(N_{\rm jets}-2)$ jets.
The 11 variables used as inputs to the neural network are summarized in Table\,\ref{tab:nnvar}. 
Comparisons of the background-dominated data and Monte Carlo generated signal events for the 11
 kinematical variables are shown in Figs.~\ref{fig:kinsel1a} to 
\ref{fig:kinsel3}.
The network is trained on same-size  samples of signal and background events
 with $6\le N_{\rm jets}\le 8$  (about 507\,000 events).
In order to model the signal we use the  {\sc pythia}\,v6.2\,\cite{Pythia}
 leading-order Monte Carlo generator with parton showering followed by a simulation of 
the CDF\, II detector. The reference top quark mass chosen for the training is 
$M_{\rm top}=175$\,GeV/$c^2$. The background is obtained from the multijet data 
events themselves, since the signal fraction at the initial stage is 
expected to be very small. 
Among the configurations investigated, the one which provides the largest expected S/B  has two hidden layers with 20 and 10 hidden nodes, respectively, and 1 output node.
 The value of the output node, $N_{out}$, is the quantity we use as discriminator between
 signal and background, as shown in  Fig.\,\ref{fig:nnout} for the $6\le N_{jets}\le 8$ sample.

\begin{table}[hbtp]
\begin{center}
\caption{Input variables to the neural network.}\label{tab:nnvar}
\begin{tabular}{cl}
\hline\hline
Variable & ~~~~~~~~~~~~~~~~~~~~Description \\
\hline
 $\sum \Et$ & Scalar sum of the transverse energies of all jets\\
 $\sum _3\Et$ &  As above, except the two highest-$E_T$ jets\\
$C$ & Centrality\\
$A$ & Aplanarity\\
$M_{2j}^{min}$ & Minimum dijet invariant mass\\
$M_{2j}^{max}$ & Maximum dijet invariant mass\\
 $M_{3j}^{min}$ & Minimum trijet invariant mass\\
$M_{3j}^{max}$ & Maximum trijet invariant mass\\
 $E_T^{\star, 1}$ & $E_T\sin^2\theta^\star$ for the highest-$E_T$ jet\\
 $E_T^{\star, 2}$ & $E_T\sin^2\theta^\star$ for the next-to-highest-$E_T$ jet\\
$\langle E_T^\star\rangle$ &  Geometric mean over the remaining jets\\
\hline\hline
\end{tabular}
\end{center}
\end{table}

\section{\label{sec:btag}Tagging $b$ quarks in the Multijet Sample}

In order to estimate the $t\bar t$ content in the event sample, we exploit the heavy flavor content of $\ttbar$ events 
using a $b$-tagging algorithm based on secondary vertex reconstruction 
as described in detail in\,\cite{vertex,massTMT}. The algorithm aims at the identification
 of jets containing a $b$-hadron state by reconstructing its decay vertex with at least two
 high-quality tracks with hits in the silicon vertex  detector. A $b$-tagged jet (``tag'', in brief) must have
 an associated secondary vertex with a displacement from the primary vertex in the transverse
 plane larger than 7.5 times the typical resolution of the vertex displacement of about 190\,$\mu$m.
Since, as shown in the next section, the background will be estimated in terms of inclusive tags rather than events, 
for the signal we do not consider the efficiency for tagging an event, but rather
the quantity to be used in the cross section calculation is 
the average number of tags per event, $n^{\rm ave}_{\rm tag}$, which is related to the cross section using  $\ttbar$ Monte Carlo calculations. The tagging efficiencies for jets coming from the fragmentation of  $b$-, $c$-, or light-flavored quarks are corrected according to the efficiency seen in  the data, by a factor $0.89\pm 0.07$ for $b$ jets
and $0.89\pm 0.14$ for $c$ jets respectively. These factors are  described
in detail in\,\cite{vertex}.

\begin{figure}[htbp]
\centering
\includegraphics[width=8.0cm]{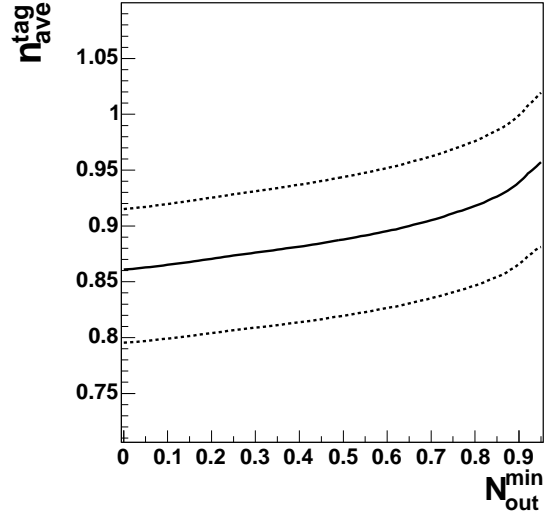}
\caption{Average number of tags, $n^{\rm ave}_{\rm tag}$, as a function of the threshold 
 $N_{out}^{min}$ for $\ttbar$ 
Monte Carlo events passing the preselection and with $6\le N_{\rm jets}\le 8$. The dashed lines represent the 1-$\sigma$ uncertainty on 
$n^{\rm ave}_{\rm tag}$.} \label{fig:btageff}
\end{figure}

We find that the average number of tags present in a $\ttbar$ event after the preselection
 depends on the choice of the lower threshold, $N_{out}^{min}$ (see Fig.\,\ref{fig:btageff}) for the neural network output $N_{out}$. 
For any value of this threshold the systematic uncertainty on $n^{\rm ave}_{\rm tag}$ is dominated by the uncertainty 
of the correction factors for tagging $b$ and $c$ jets.

\section{\label{sec:bkg}Background Estimate}
\begin{figure}[htbp]
\centering
\includegraphics[width=8.0cm]{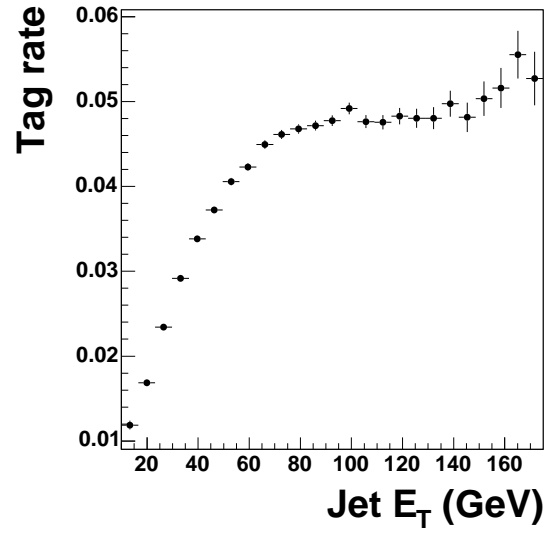}
\hfill
\includegraphics[width=8.0cm]{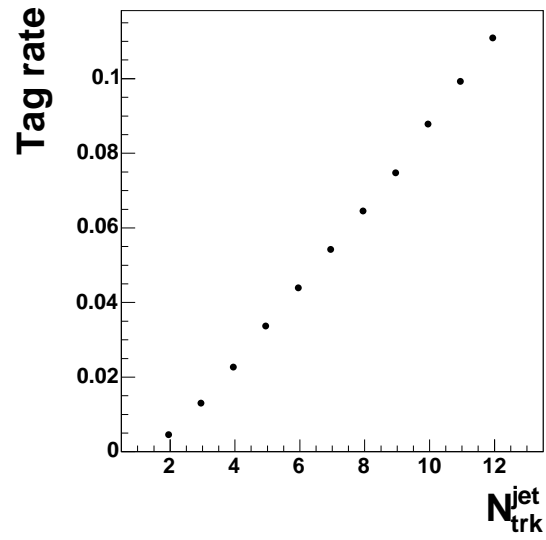}
\hfill
\includegraphics[width=8.0cm]{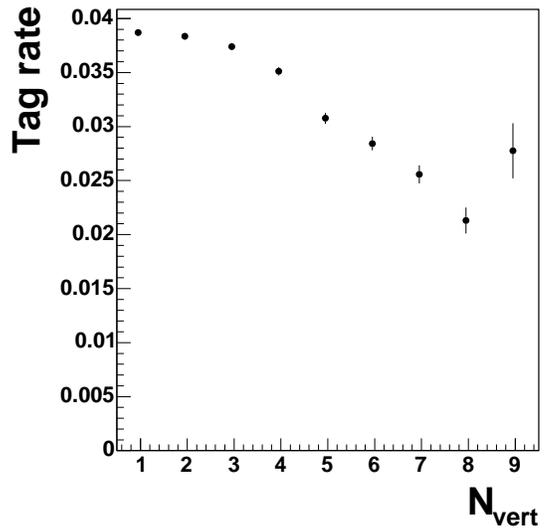}
\hfill
\caption{Tag rate for fiducial jets as a function of jet $\Et$, $N_{\rm trk}^{jet}$ and $N_{\rm vert}$, as  measured in the control sample with exactly four jets.} \label{fig:tagrate}
\end{figure}

The background for the  $\ttbar$ multijet final state comes  mainly from QCD production of 
heavy-quark pairs ($b\bar b$ and $c\bar c$) and false tags in light-quark jets. 
Other standard model processes such as $W/Z$+jets have  a smaller
 production cross section and small acceptance due to the selection cuts. 
Due to the current uncertainties on the Monte Carlo generation of QCD multijet events,  we estimate the
background  from the data themselves.
Given the theoretical uncertainties on the production cross section for  
 events generated with N partons at tree level, a more accurate background estimate is obtained from the data themselves rather than from 
theoretical prediction of cross section and simulations.
The tag rate per jet is defined as the probability of tagging a jet whose tracks are reconstructed in the vertex detector. This rate is extracted from events depleted in $\ttbar$ signal, and is used as an estimate of the rate expected in events with 
different jet multiplicity. 
These depleted events, with exactly four jets passing the preselection (``control sample''), 
have S/B $\approx 1/3600$. 
This method intrinsically provides an inclusive estimate in terms of number of tags  rather than number of tagged events.
The tag rate per jet is evaluated in this control sample and is parameterized in terms 
of variables sensitive to both the tagging efficiency for true heavy-flavored objects and the rate of false tags.
These variables are jet $\Et$, the number of tracks reconstructed in the vertex detector 
and associated to the jet, $N_{\rm trk}^{jet}$, and the number of primary vertices in
 the event, $N_{\rm vert}$. The tag rates per jet as a function of these variables
 are shown in Fig.\,\ref{fig:tagrate} for jets with at least two tracks within the vertex detector acceptance (``fiducial'' jets).

%%%%%%%%%%%%%%%%%
%%%%%%%%%%%%%%%%%%

\begin{figure}[htbp]
\begin{center}
\includegraphics[width=8.0cm]{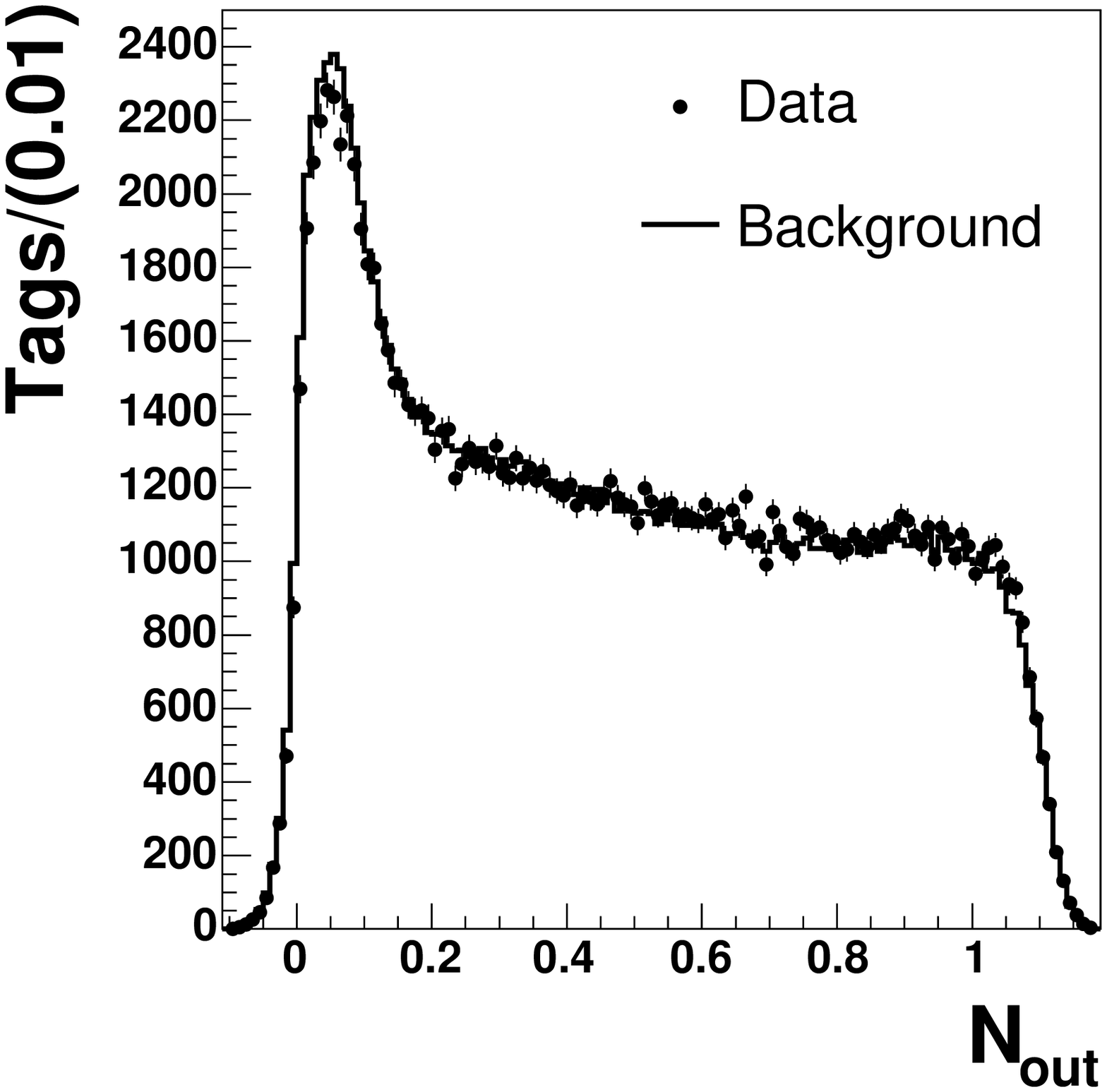}
\caption{Neural network output, $N_{out}$, distribution for tags in the data events with 4 jets, compared with 
the estimate from the tag rate parametrization. Events with multiple tags have  multiple entries.} \label{fig:nntag4}
\end{center}
\end{figure}

\begin{figure}[htbp]
\begin{center}
\includegraphics[width=8.0cm]{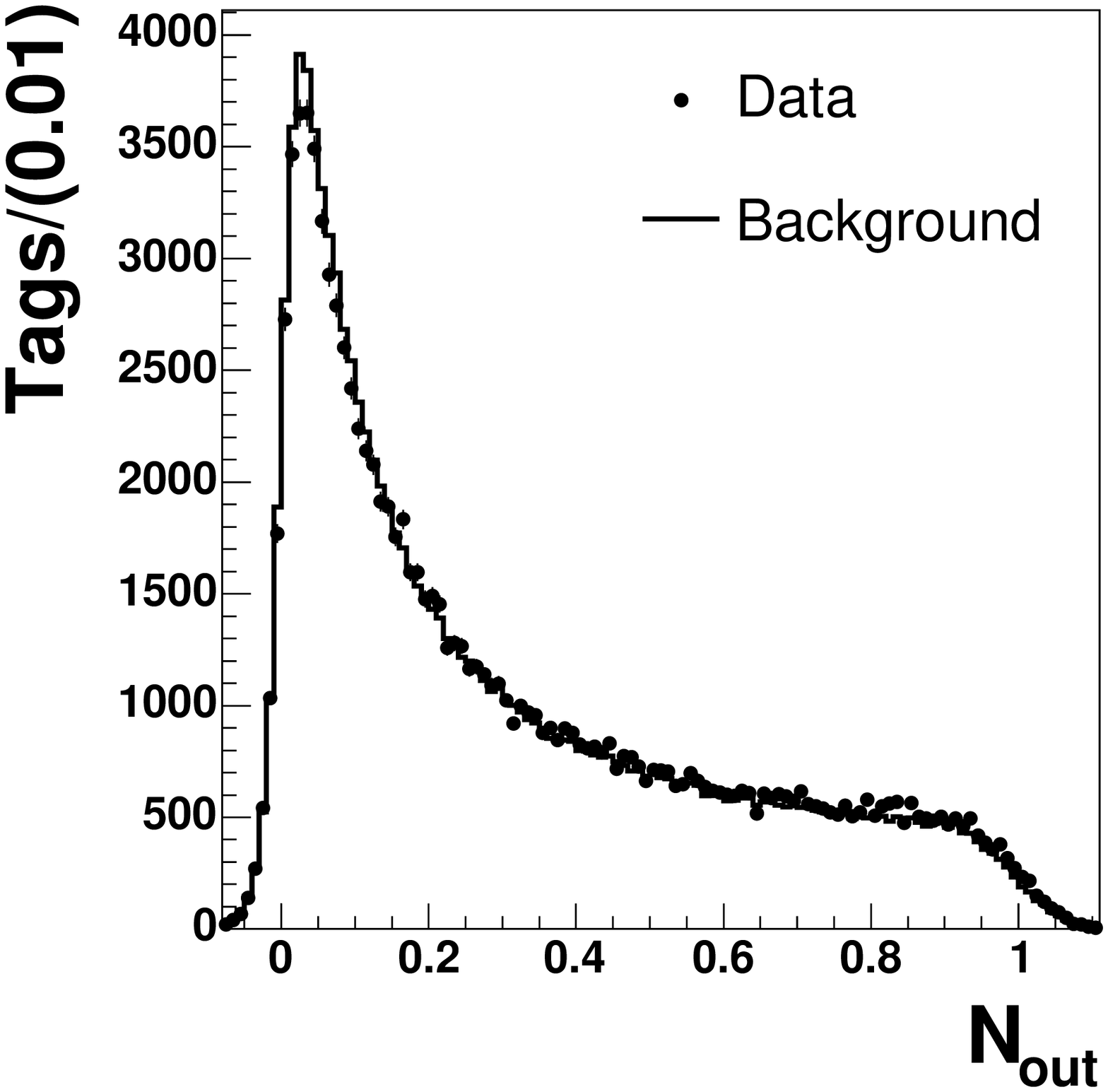}
\caption{Neural network output, $N_{out}$, distribution for tags in the  data events with 5 jets, compared with 
the estimate from the tag rate parametrization. Events with multiple tags have  multiple entries.} \label{fig:nntag5}
\end{center}
\end{figure}

\begin{figure}[htbp]
\begin{center}
\includegraphics[width=8.0cm]{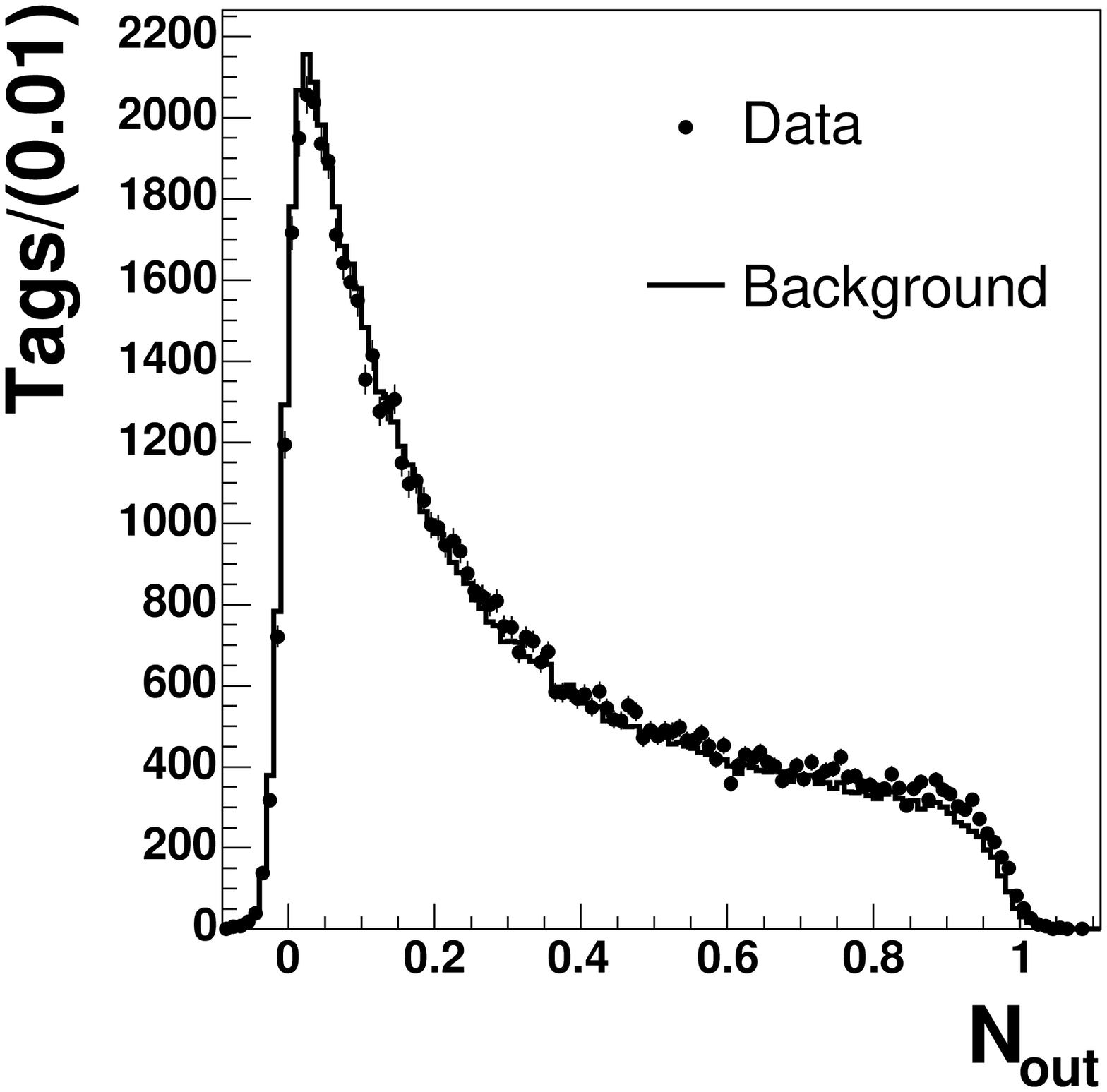}
\caption{Neural network output, $N_{out}$, distribution for tags in the data events with 6 to 8 jets, compared
 with the estimate from the tag rate parametrization. Events with multiple tags have  multiple entries.} \label{fig:nntag6}
\end{center}
\end{figure}

The tag rate estimates the probability that a given fiducial jet
 in the signal sample is tagged. Summing this probability over all fiducial jets, we obtain 
the expected number of tags from non-signal processes, that is QCD heavy- and
 light-flavored production altogether. 
Before the neural-network-based  kinematical selection, the multijet sample is composed 
essentially of background events. 
The accuracy of our modeling of the background processes, assuming all jets to be uncorrelated,
is shown in Figs.~\ref{fig:nntag4}, \ref{fig:nntag5}, and \ref{fig:nntag6} where we compare the $N_{out}$ distributions for tags
 in the data to the expected background in events with 4, 5 and 6 or more jets. The disagreement 
between the total number of observed and expected tags, possibly due to the presence 
of a small sample of $t\bar t$ events, amounts to no more than 0.8\%.
This small discrepancy  observed at high jet multiplicity is accounted for as a systematic
 uncertainty on the background estimate due to the different jet multiplicity in the signal and control samples. 
%%%%%%%%%%%%%%%%%
%%%%%%%%%%%%%%%%%
%%%%%%%%%%%%%%%%%
%%%%%%%%%%%%%%%%%%
The kinematical selection cutting on $N_{out}$ also changes the event characteristics with respect to those found in
 the 
sample with exactly four jets, where the parameterization has been derived. This selection
 modifies the jet-$E_T$ and $\eta$ spectra so that the average tag rate per event for jets
 from QCD background becomes higher. 
However, the parameterization of the tag rate in terms of properties of the jet ($E_T$ and 
$N_{trk}^{jet}$) is shown to describe well this increase.  
Residual biases due to the neural network selection are treated as systematic uncertainties 
on the background prediction considering, in the control sample with four jets, subsets of 
events with $N_{out}\ge N_{out}^{min}$ with $N_{out}^{min}$ in the range $0.8-1.0$.
 In this case the disagreement between the total number of observed tags and the expected
 background is no more than about 2.4\%, due to the neural network kinematical selection.
 The total systematic uncertainty expected on the background estimate amounts then to
 2.5\%, the quadrature sum of the  two uncertainties.
The contributions from running conditions, such as instantaneous luminosity and detector 
configuration, have been studied and found to be negligible. 

\section{\label{sec:optim}Optimization of the kinematical selection and its efficiency}

\begin{figure}[htbp]
\begin{center}
\includegraphics[width=8.0cm]{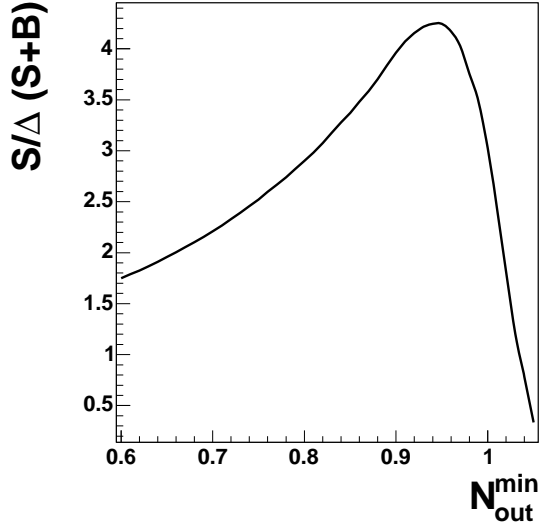}
\caption{Signal significance as a function of the neural network threshold
 $N_{out}^{min}$.} \label{fig:optim}
\end{center}
\end{figure}

In order to obtain the most precise cross section measurement, the neural-network-based  kinematical selection is optimized, 
in the sample  $6\le N_{\rm jets}\le 8$, for the maximum signal 
significance for $\ttbar$ events, defined as the ratio between the expected signal 
(assuming the theoretical production cross section of 6.7 pb) and the total uncertainty on
 the sum of signal and background, where both statistical and systematic uncertainties are
 considered. Since we have an accurate background prediction only after the $b$-tagging requirement, 
the optimization refers to tagged $t \bar t$ events and the expected background.
The cut on the neural network output which provides the maximum signal significance is
$N_{out}\ge 0.94$, as can be seen in Fig.\,\ref{fig:optim}.
Such a selection yields, before $b$-tagging, 4205 candidate events in the data with an efficiency of 
$4.8\%$
for the $\ttbar$ signal and with S/B $\sim 1/12$. 

The effect of the selection on $\ttbar$ events and on the data is summarized in 
Table~\ref{tab:kineff}. The relative contribution from the leptonic channels after 
all the cuts is small, about 3\%. A summary of the data as a function of jet 
multiplicity is shown in Table~\ref{tab:jetmult}. 

%%%%%%%%%%%%%%%%%%

\begin{table}[hbtp]
\begin{center}
\caption{Efficiency of the kinematical selection measured from {\sc pythia} $\ttbar$ 
Monte Carlo simulation for $M_{\rm top}$=175\,GeV/$c^2$ and number of events selected in the data (before requiring $b$-tagging).}\label{tab:kineff}
\begin{tabular}{lcc}
\hline\hline
Requirement & Efficiency (\%) & Data \\
\hline
Trigger & 58 & 4340143\\
Preselection & 38 & 3480768\\
$6\le N_{\rm jets}\le 8$ & 20 & 506567\\
$N_{out}\ge 0.94$ & 4.8 & 4205\\
\hline\hline
\end{tabular}
\end{center}
\end{table}
%%%%%%%%%%%%%%%%

%%%%%%%%%%%
\begin{table}[hbtp]
\begin{center}
\caption{The number of events in data before and after the neural network selection, 
$N_{out}\ge 0.94$, for different jet multiplicities (before requiring $b$-tagging).}
\label{tab:jetmult}
\begin{tabular}{c c c}
\hline\hline
Jet & Events before & Events after \\
Multiplicity & selection & selection \\
\hline
4 & 1341622 & 118657\\
5 & 917999  & 16157 \\
6 & 372091  & 2575 \\
7 & 109295   & 1069 \\
8 & 25181   & 561 \\
\hline\hline
\end{tabular}
\end{center}
\end{table}
%%%%%%%%%

The systematic uncertainties affecting the $\ttbar$ 
%efficiency are summarized in 
production cross section are summarized in
Table~\ref{tab:kineffsyst}, with a total relative contribution of 17\% on the kinematical selection efficiency. 
The uncertainty of 16.3\% arising from the jet energy scale is dominant, since this 
analysis requires the presence of a large number of jets in the event which are used
 to build the set of kinematical variables employed in the selection. This uncertainty is evaluated by the shift in
signal acceptance observed by changing the jet energy corrections within their uncertainties.
Less relevant sources of uncertainty are associated with  different Monte Carlo 
hadronization schemes ($1.1\%$), increased and decreased initial and final state radiation (ISR and FSR) ($2.9\%$), and 
the variation of parton distribution functions (PDFs) within their uncertainties ($1.4\%$). 
A detailed description of the procedure used to estimate these uncertainties can be found in\,\cite{ljetsKIN}. The difference 
in the amount of multiple beam interactions present in the data events and in the Monte Carlo simulation is also accounted for 
($2.5\%$).

%%%%%%%%%%%%%%%%
\begin{table}[hbtp]
\begin{center}
\caption{Relative systematic uncertainties on the signal efficiency, and other uncertainties related to the cross section.}\label{tab:kineffsyst}
\begin{tabular}{lc}
\hline\hline
Source & Uncertainty (\%) \\
\hline
Energy Scale & 16.3 \\
Parton Distribution Functions        & 1.4 \\
Initial/Final State Radiation     & 2.9 \\ 
Monte Carlo Modeling & 1.1 \\
Multiple interactions & 2.5\\
\hline
%Total        & 17.0 \\
Average number of tags & 7.4\\
\hline
Estimated background & 2.5\\
\hline
Integrated luminosity & 6.0\\
\hline\hline
\end{tabular}
\end{center}
\end{table}
%%%%%%%%%%%%%%%%

After the kinematical selection with $N_{out}\ge 0.94$, the $b$-tagging selects 1020 events
 with 1233 candidate tags.  The estimated background amounts to $937\pm 30$ tags, while for  $\ttbar$ 
events we expect an average number of tags $n^{\rm ave}_{\rm tag}=0.95\pm 
0.07$.  Since the background estimate is obtained from all the events passing the selection
 before tagging, we need to subtract the contribution due to the $t\bar t$ events, as obtained
from the excess of candidates with respect to the background, divided by $n^{\rm ave}_{\rm tag}$\,\cite{cdftop}. This contribution is then subtracted
from the number of events before tagging to obtain, iteratively, a new background estimate.
After this correction, the number of tags expected from background sources is reduced to 
 $846\pm 37$ tags, where the increased uncertainty accounts for the uncertainty on  $n^{\rm ave}_{\rm tag}$.

%%%%%%%%%%%%%%%%%
\begin{figure}[htbp]
\begin{center}
\includegraphics[width=8.0cm]{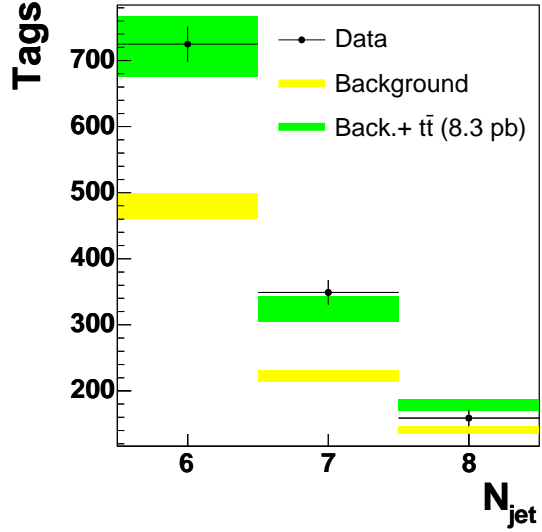}
\caption{Number of candidate tags as a function of the jet multiplicity for the data after the kinematical 
selection in the signal region, compared with the expected background. The $\ttbar$ expectation 
is based on the measured cross section of 8.3\,pb. See Table\,\ref{tab:tagskin}.} \label{fig:tagskin}
\end{center}
\end{figure}
%%%%%%%%%%%%%%%%%%

\begin{table*}[hbtp]
\begin{center}
\caption{Observed number of tags and expected background and signal after the kinematical selection $N_{out}\ge 0.94$. The corrected background accounts for the presence of $t \bar t$ events before tagging. The uncertainties correspond to the quadrature sum of statistical and systematic uncertainties. See Fig.\,\ref{fig:tagskin}.}\label{tab:tagskin}
\begin{tabular}{c cc|ccc}
\hline\hline
Jet Multiplicity & 4 & 5 & 6 & 7 & 8 \\
\hline
Background & $16060\pm 575$ & $2750\pm 92$& $536\pm 17$& $255\pm 8$ & $146\pm 5$ \\
Corrected Background & $15961\pm 677$ & $2653\pm 112$& $481\pm 20$ & $223\pm 10$ & $142\pm 7$\\
$\ttbar$ ($\sigma_{t \bar t}=8.3$~pb) & $120\pm 20$ & $266\pm 45$& $242\pm 41$& $101\pm 17$&$38\pm 7$\\
Background + $t \bar t$ & $16081\pm 677$ & $2919\pm 121$& $723\pm 46$ & $324\pm 20$ & $180\pm 10$\\
\hline
Data & 16555 & 3139 & 725 & 349 & 159\\
\hline\hline
\end{tabular}
\end{center}
\end{table*}
%%%%%%%%%%%%%%%%

\section{\label{sec:xs}Cross Section Measurement}
\begin{figure}[htbp]
\begin{center}
\includegraphics[width=8.0cm]{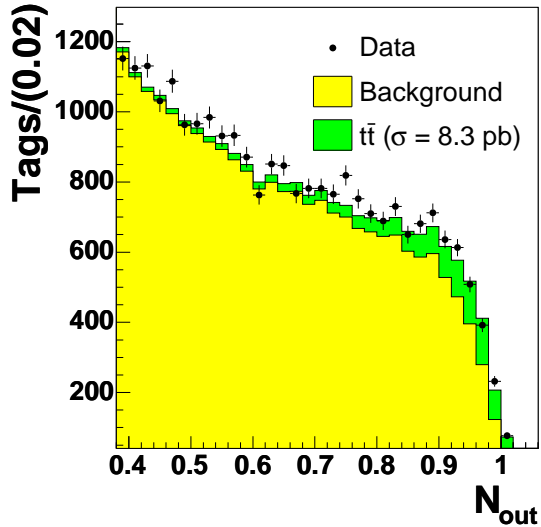}
\caption{$N_{out}$ distribution for the tags in the data, compared with the expected 
background. The $\ttbar$ expectation is based on the measured cross section of 8.3\,pb.} \label{fig:nnalltag}
\end{center}
\end{figure}

In the signal region $6\le N_{\rm jets}\le 8$,  the excess of observed candidate tags in the data over the background is ascribed to $\ttbar$ production. A measurement of the 
cross section can be extracted from the acceptance and the background estimate: 
\begin{equation}\label{eq_xsec}
\sigma_{\ttbar}=\frac{N_{\rm obs}-N_{\rm bkg}}{\epsilon_{\rm kin} \times n^{\rm ave}_{\rm tag}\times {\cal{L}}_{\rm int}}
\end{equation}
where $N_{\rm obs}=1233$ and $N_{\rm bkg}=846\pm 37$ are the number of total observed and 
background tags, respectively, in the signal region $6\le N_{\rm jets}\le 8$,
 $\epsilon_{\rm kin}=4.8\pm~0.8\%$ is the signal kinematical selection efficiency, 
$n^{\rm ave}_{\rm tag}=0.95\pm 0.07$ is the average number of tags in $\ttbar$ events, 
and ${\cal L}_{\rm int}=1.02\pm 0.06$\,fb$^{-1}$ is the integrated luminosity of the data sample.
Given the nature of our background estimate, in the above formula we use tagged
 jets instead of tagged events. We have verified however that this
 does not introduce any bias in the cross section, but leads to a small 
underestimate of the statistical uncertainty, due to the fact that some events have two or more tags. The statistical uncertainty is inflated 
appropriately ($\approx +20\%$), as determined from  a set of Monte Carlo simulations where 
the number of expected signal and background events with 0, 1, 2 or more tags are fluctuated according to Poisson distributions.
The measured value of the $\ttbar$ cross section is
$\sigma_{\ttbar}=8.3\pm 1.0({\rm stat.})^{+2.0}_{-1.5}({\rm syst.})\pm 0.5({\rm lumi.})$\,pb
 for a top quark mass of 175\,GeV/$c^2$. 
In Fig.\,\ref{fig:tagskin} the distribution of the number of observed tags and expected
 background is compared in the signal region  to the $\ttbar$ signal expectation assuming
 the production cross section measured in this analysis. A good agreement is 
observed for all jet multiplicities after the kinematical selection $N_{out}\ge 0.94$, as can be seen  in Table~\ref{tab:tagskin}.
In addition we show 
in Fig.\,\ref{fig:nnalltag} that the measured cross section is in agreement with the data over a wide $N_{out}$ range.

%%%%%%%%%%%%%%%%%
%%%%%%%%%%%%%%%%%%

\section{\label{sec:chi2} Mass reconstruction}

Having evaluated the $t \bar t$ production cross section in the preceding section, 
we proceed to measure the top quark mass from the same candidate events with a constrained-fitting technique. 
Since the kinematical selection described above is designed to have the best accuracy on the cross section measurement, we allow here for a different cut on $N_{out}$ in order to find the value
which provides the best accuracy  for the mass measurement.
We search in simulated $t\bar t$ events the $N_{out}^{min}$ value which provides the smallest 
 statistical uncertainty expected on the mass measurement\footnote{The systematic uncertainty expected 
on the mass measurement is rather insensitive to the neural net cut and has been neglected in the choice of the best performing value  of  $N_{out}^{min}$.}.

For each event we determine a reconstructed top quark mass, $m_t^{reco}$, from the four-momenta of the six highest-$E_T$ jets.
This mass is entered into a distribution which is used to determine the most likely top quark mass from the data sample
(see Section\,\ref{sec:mass}).

\subsection{\label{sec:kinfit} Kinematical fitter}

Sixteen equations can be considered to connect the four-momenta of the two top quarks and the six final state particles according to the $t \bar t \rightarrow b \bar b \, W^+ W^- \rightarrow b \bar b \, q_1 \bar q_2 \, q_3 \bar q_4$ hypothesis:
\begin{eqnarray}
 p_t^\mu        & = &  p_{W^+}^\mu + p_b^\mu       \\
 p_{\bar t}^\mu & = &  p_{W^-}^\mu + p_{\bar b}^\mu \\
 p_{W^-}^\mu    & = &  p_{q_1}^\mu + p_{\bar q_2}^\mu \\
 p_{W^+}^\mu    & = &  p_{q_3}^\mu + p_{\bar q_4}^\mu 
\end{eqnarray} 
\begin{displaymath}
(\mu   =  0, 1,2,3)
\end{displaymath}

There are 13 unknown quantities, i.e., the unknown top 
quark mass and the three-momenta of the top quarks and of the $W$ bosons, so the kinematics of the event is overconstrained.
Only the six highest-$E_T$ jets are used as inputs to the kinematical fitter, 
resulting in 90 possible permutations of two jet doublets giving a $W$ and of two jet triplets giving the top quarks.
Only events with at least one $b$ tag among the six  highest-$E_T$ jets are used in this analysis, with the association of the tagged jet to a $b$ quark,
reducing the number of possible jet-to-parton permutations to 30.
We construct the $\chi^2$ function
\begin{displaymath}
 \chi^2 = \frac{(m_{jj_1}-m_W)^2}{\Gamma^2_W} + \frac{(m_{jj_2}-m_W)^2}{\Gamma^2_W} + \frac{(m_{jjj_1}-m_t^{reco})^2}
{\Gamma^2_t} 
\end{displaymath}
\begin{equation}
 + \frac{(m_{jjj_2}-m_t^{reco})^2}{\Gamma^2_t} + \Sigma^6_{i=1} \frac{(p^{fit}_{T,i}-p^{meas}_{T,i})^2}{\sigma^2_i}, 
\end{equation}

where $m_{jj_{1,2}}$ are the invariant masses of the dijet systems, $m_{jjj_{1,2}}$ are the invariant masses of the trijet 
systems, $\Gamma_W=2.1$ GeV/c$^2$ is the measured natural width of the $W$\,\cite{widW},  and
$\Gamma_t$, fixed to $1.5$ GeV/c$^2$, is the assumed natural width of the top 
quark\,\cite{widt}. 
The measured jet transverse energies, $p^{meas}_{T,i}$ are free to vary within their known resolution, $\sigma_i$.
The $\chi^2$ is minimized with respect to the 7 free parameters ($m_t^{reco}$, and the 6 jets 
transverse momenta $p^{fit}_{T,i}$) for each of the 30 permutations of jets with final state partons. 
The permutation with the lowest $\chi^2$ is selected and  a distribution (``template'') of $m_t^{reco}$ is then formed 
to be used for the determination of the true top quark mass.

In order to reconstruct a data-driven background template we apply the kinematical fitter
to the sample of events passing the neural network selection, but before the requirement of identified $b$ jets. 
Within an event the fit is performed once for each fiducial jet, assuming it is a $b$-quark. The resulting value of $m_t^{reco}$
 enters the template with a weight given by the tag rate associated with the fiducial jet.
The integral of the $m_t^{reco}$ distribution is the sum of all weights and  represents the expected number of background tags.
This procedure does not allow a separation between the background expected for events with 1 or 2 tags, but treats them together. 

\begin{figure} [htbp]
\centering
\includegraphics[width=8.0cm]{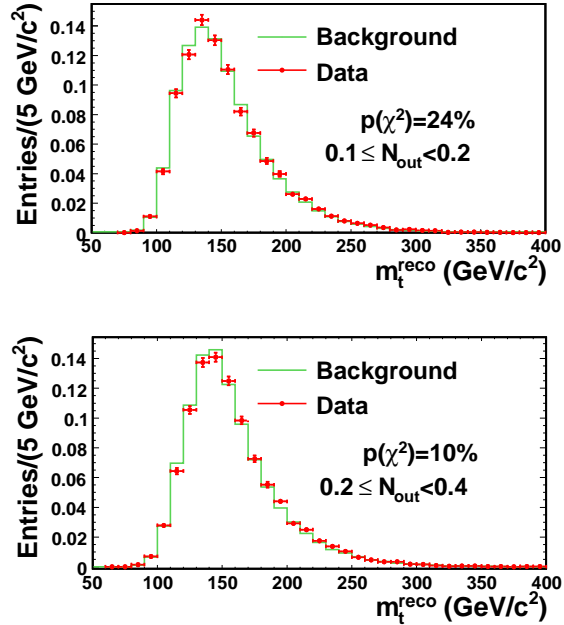}
\caption{Comparison of the expected versus observed $m_t^{reco}$. Control regions $0.1\le N_{out} < 0.2$ (top) and $0.2\le N_{out} < 0.4$ (bottom). The $\chi^2$ probability is indicated in the legend.} 
\label{fig:fitnn1}
\end{figure}

\begin{figure} [htbp]
\centering
\includegraphics[width=8.0cm]{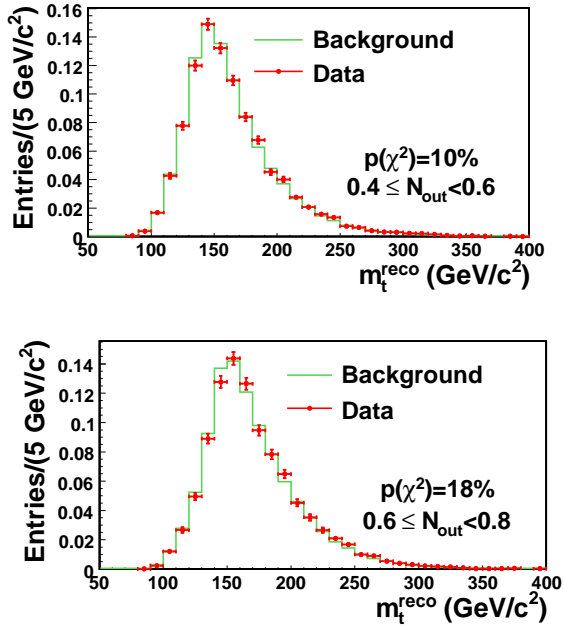}
\caption{Comparison of the expected versus observed $m_t^{reco}$. Control regions $0.4\le N_{out} < 0.6$ (top) and $0.6\le N_{out} < 0.8$ (bottom). The $\chi^2$ probability is indicated in the legend.} 
\label{fig:fitnn2}
\end{figure}

We follow the same approach for the data and the signal simulation, so the fit is performed for each association of tagged jets with one or the other of the two $b$ quarks; events with more than 1 tag contribute then with multiple entries.
A control sample where we expect the signal to be present in fractions of few percent or below, 
is defined by $0.1\le N_{out} \le 0.8$ and is  further subdivided in 4 subregions, 
to check the goodness of the background modeling. As shown in Figs.\,\ref{fig:fitnn1} and \,\ref{fig:fitnn2}, the background
is found to describe well the $m_t^{reco}$ distributions in all the subsamples.


\section{\label{sec:like}Likelihood fit}

The technique described above gives a distribution of $m_t^{reco}$ (one for each tag in the 
event) in the corresponding data sample, which is a mixture of signal and background. In order to measure the top quark 
mass, we compare the $m_t^{reco}$ distribution from the data to the signal and background templates. 
From the templates we first derive probability density functions (p.d.f.'s) and then perform an unbinned likelihood fit 
to determine the value of true top quark mass, $\mtop$, that best describes the data.

\subsection{The Likelihood Function}

The p.d.f.'s for the $m_t^{reco}$ distributions are parametrized in order to have  
a functional form which varies smoothly with $\mtop$.
For the signal, we use $t \bar t$ Monte Carlo events generated with {\sc herwig} v6.508 \cite{Herwig} with top quark masses ranging from 150 to 200 GeV/c$^2$ in $2.5$ GeV/c$^2$ increments. 
The function for the signal p.d.f., $P_{sig}(m|M_{\rm top})$, represents the probability to obtain a value $m$ for $m_t^{reco}$, given a true top quark mass $M_{\rm top}$ in a $t\bar t$ event. The form used is a sum of a Gamma distribution, 
chosen to describe the reconstructed top quark mass in case of incorrect jet-parton assignments, and two Gaussian distributions, 
which model the core of the distribution. Its explicit expression is 
\begin{displaymath}
 P_{sig}(m|M_{\rm top}) = \delta_7 \cdot \frac{\delta^{1+\delta_1}_2}{\Gamma(1+\delta_1)} \cdot (m- \delta_0)^{\delta_1} \cdot \exp(-\delta_2(m-\delta_0)) 
\end{displaymath}
\begin{displaymath}
 + \delta_8 \cdot \frac{1}{\sqrt{2\pi} \delta_4} \cdot \exp\left(\frac{-(m-\delta_3)^2}{2\delta^2_4}\right) 
\end{displaymath}
\begin{equation}
 + (1- \delta_7 - \delta_8) \cdot \frac{1}{\sqrt{2\pi} \delta_6} \cdot \exp\left(\frac{-(m-\delta_5)^2}{2 \delta^2_6}\right)        
\end{equation}

where each parameter $\delta_i$ is linearly dependent on $\mtop$,
\begin{equation}
 \delta_i = \alpha_i + \beta_{i} \cdot (M_{\rm top}-175) ~~~~~~~~ (i=0,1,...8)  
\end{equation}

so the total number of parameters used is 18. 

\begin{figure}[htbp]
\centering
\includegraphics[width=8.0cm]{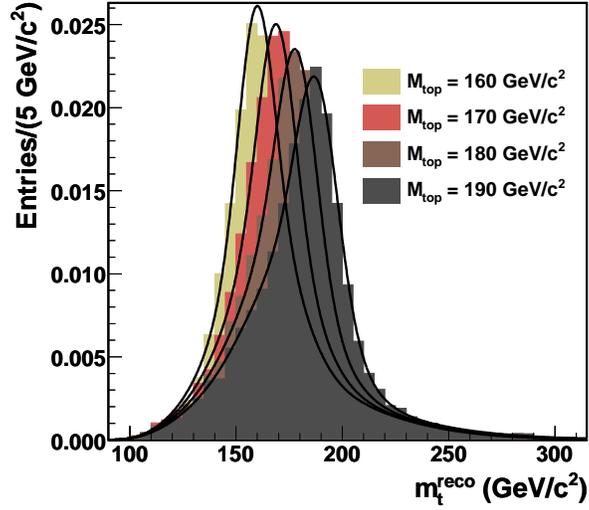}
\caption{Four $m_t^{reco}$ signal templates with their p.d.f. overlaid.} 
\label{fig:sigmass}
\end{figure}

Fig.\,\ref{fig:sigmass}. shows some of the signal templates along with 
their parametrized  p.d.f.'s. \\

\begin{figure}
\begin{center}
\includegraphics[width=8.0cm]{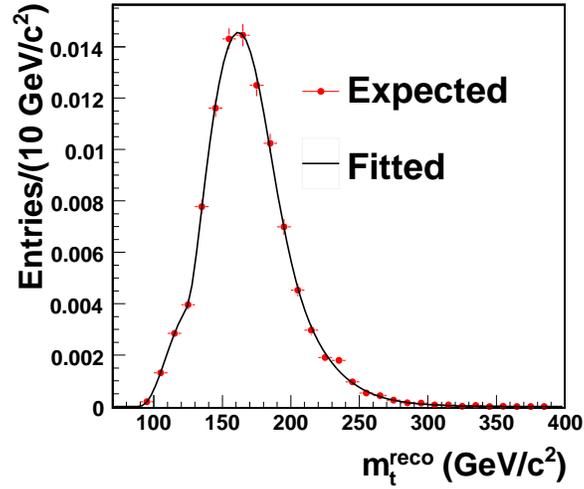}
\caption{$m_t^{reco}$ background template with its p.d.f. overlaid.} 
\label{fig:fondo}
\end{center}
\end{figure}

The background reconstructed mass is computed as described in the previous section.
The integral of this distribution is the predicted amount of background tags, which is 
corrected for the presence of tags expected from $t\bar t$ events using the iterative technique described in Section \ref{sec:optim}. 
In order to account for the shape of the signal mass distribution, the correction is made by substracting from the background distribution the residual $t\bar t$
contamination distributed in mass as expected for  $\mtop = 175$ GeV/c$^2$. 
The systematic uncertainty associated with this procedure is estimated and reported in the next section.
The resulting background template is parametrized with two Gamma distributions and one Gaussian distribution.
The background distribution does not  depend on the top quark mass.
The resulting p.d.f., $ P_{bkd}(m)$, is

\begin{displaymath}
 P_{bkd}(m) = \delta_8 \cdot \frac{\delta^{1+\delta_1}_2}{\Gamma(1+\delta_1)} \cdot (m- \delta_0)^{\delta_1} \cdot \exp(-\delta_2(m-\delta_0)) 
\end{displaymath}
\begin{displaymath}
 + \delta_9 \cdot \frac{\delta^{1+\delta_4}_5}{\Gamma(1+\delta_4)} \cdot (m- \delta_3)^{\delta_4} \cdot \exp(-\delta_5(m-\delta_3)) 
\end{displaymath}
\begin{equation}
 + (1-\delta_8-\delta_9) \cdot \frac{1}{\sqrt{2 \pi} \delta_7} \exp\left(\frac{-(m-\delta_6)^2}{2 \delta^2_7}\right)       
\end{equation}

and is plotted in Fig.\,\ref{fig:fondo}.

The likelihood function, ${\cal L}$,  is constructed by assuming that the data are described by 
an admixture of background and $t \bar t$ events with a 
certain top quark mass.
The function is obtained by multiplication of three terms. 
The first two terms constrain the number of background tags, $n_b$, to the expectation, $n_b^{exp}$, and the number of signal and background tags, $n_s +  n_b$, to be equal to the number observed in the data, $N$.
In the third term the signal and background probabilities are assigned by comparing the measured 
value $m_i$ of $m_t^{reco}$ from the data with the parameterized signal and 
background p.d.f.'s, $P_{sig}$ and $P_{bkd}$. 

\begin{displaymath}
 {\cal L} = \exp\left(-\frac{(n_{b}-n_b^{exp})^2}{2\sigma^2_{n_b}}\right) \times  \exp\left(-\frac{(n_s+n_b-N)^2}{2\sigma^2_N}\right)
\end{displaymath}
\begin{equation}
 \times \prod_{i=1}^N \frac{n_s \cdot P_{sig}(m_i|M_{\rm top}) + n_b \cdot P_{bkd}(m_i)}{n_s+n_b} 
\end{equation}

where $\sigma_{n_b} $ is the expected uncertainty  on the corrected background, 
and $\sigma_{N} = \sqrt{N}$ is the expected uncertainty  on the total number of observed tags, $N$.
In order to facilitate the computation, we minimize the negative logarithm of the likelihood, $-ln{\cal L}$, 
instead of maximizing the likelihood itself. The minimization is performed with respect to the three 
free parameters, $n_s, n_b$ and $M_{top}$. The statistical uncertainty on the top quark mass is  taken from the points where the
$-ln{\cal L}$ changes by 0.5 units from its minimum. 

\subsection{\label{sec:sanity}Verification and Calibration of the Method}

\begin{figure}
\begin{center}
\includegraphics[width=8.0cm]{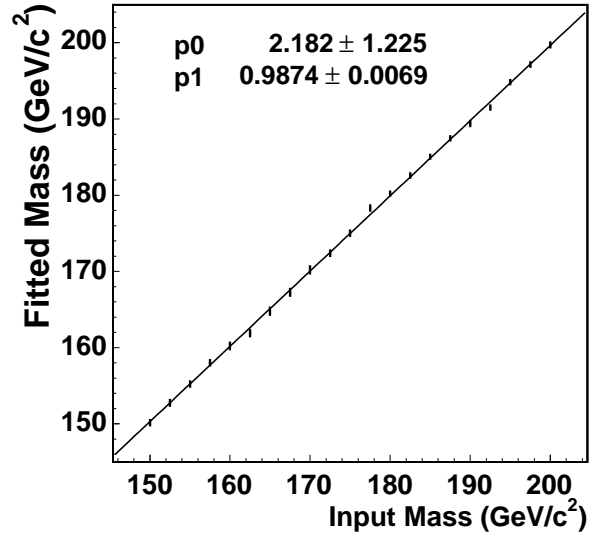}
\caption{Fitted mass as a function of the input top quark mass using pseudo-experiments. The fit slope is consistent 
with 1. The coefficients of the fitted straight line are indicated in the legend.}
\label{fig:linear}
\end{center}
\end{figure}

\begin{figure}
\begin{center}
\includegraphics[width=8.0cm]{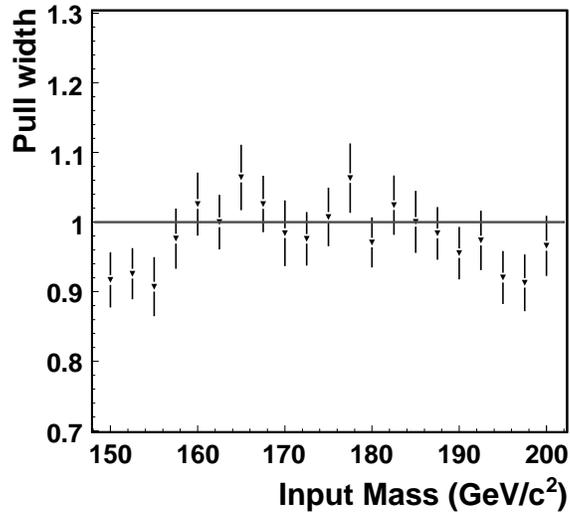}
\caption{Width of the pull distributions, for different top quark input masses.} 
\label{fig:mewid}
\end{center}
\end{figure}

We investigate for possible biases in the top quark mass measurement which can be introduced by our method,
and quantify its statistical power before performing the actual measurement on the data sample.
We run Monte Carlo simulations of the experiment (``pseudo-experiments''), generated with  the true 
 top quark mass ranging from 150 to 200 GeV/c$^2$, and then extract the predicted amount of  $m_t^{reco}$ values 
from the signal and background templates. We fix the total number of tags in each pseudo-experiment to be the same as observed in the data.
Then we take our background expectation of tags passing the cuts, fluctuate it according to Gaussian statistics, and get the number of signal tags as the difference between $N$ and the fluctuated background.
We perform the measurement on many different sets of Monte Carlo events in the pseudo-experiment (``pseudo-events''), 
and plot the fitted top 
quark mass with respect to the input mass in Fig.\,\ref{fig:linear}, seeing no systematic bias.
We define the ``pull'' of the fit variable to be the 
 deviation of the fitted mass from the true value in the pseudo-experiment, divided by the measurement uncertainty determined in the fit.
The pull distribution is fitted with a Gaussian and its width (``pull width'') indicates the accuracy of the 
uncertainties obtained from the fit (see Fig.\,\ref{fig:mewid}).

The sensitivity check is performed over a range of thresholds for both $N_{out}$ and $\chi^2$, to achieve the smallest  expected 
statistical uncertainty. The $N_{out}^{min}$ cut is varied between 0.88 and 0.96 while the 
upper threshold, $\chi^2_{max}$, varies between 30
and 6. The cuts for which we expect the smallest statistical uncertainty on the mass 
measurement  are $N_{out} \ge 0.91$ and $\chi^2 \le 16$.
The  $\chi^2$ cut improves the background rejection so that we need to apply a threshold on $N_{out}$ lower 
than that used for the cross section measurement.

\section{\label{sec:masssyst}Systematic uncertainties on the top quark mass}
Various sources of systematic uncertainty affect the top quark mass measurement.
 Systematic effects arise
from mismodeling in the simulation of the detector response to jets, and from uncertainties in the simulation of 
the $t \bar t$ signal. The evaluation of the mass shift due to each source of systematic uncertainty consists in 
generating pseudo-experiments varying the effects from  each possible source by $\pm 1 \sigma$, building new templates out of the varied sample
and determining how much the fitted top quark mass shifts.
We take as an estimate of the respective uncertainty the resulting half-difference between the 
extreme values of the measured mass. If the shift is smaller than the statistical uncertainty 
on the estimate itself, we assign the latter 
as the systematic uncertainty. 

The main contribution to the systematic uncertainty stems from the residual uncertainty 
on the jet energy after it is corrected for the
known effects \cite{JESNIM}. These include calorimeter non-linearity in response to single hadronic particles, energy
loss into  non-instrumented calorimeter regions, energy added to the jet by secondary interactions and the underlying 
event, and  energy lost outside the jet cone. 
We calculate the systematic uncertainty originating from each of these sources varying each corrected jet energy in 
the simulation by the corresponding uncertainty and performing pseudo-experiments with the modified resulting templates, and 
finally adding the observed mass shifts in quadrature to quote a total systematic due to the jet energy scale.
Since the jet energy corrections are derived on data samples deprived of heavy quarks, we add an additional uncertainty 
of the order of 0.6\%
evaluated  considering the different fragmentation properties of $b$ quarks\,\cite{ljetsSVX}. 
As done for the generic jet energy scale uncertainties, we perform pseudo-experiments where we vary the $b$-jet energy scale accordingly and use half 
the variation  
in the fitted top quark mass as the $b$-jet energy scale uncertainty.
\par
Many sources of systematic effects arise from uncertainty in the Monte Carlo modeling of
 the hard interaction. {\sc pythia} and {\sc herwig} 
generators differ in their hadronization  schemes and in their description of the underlying event and multiple interactions. 
A corresponding systematic uncertainty is evaluated drawing top quark reconstructed masses from 
{\sc pythia}-generated events and comparing the resulting mass distributions with the 
template constructed using {\sc herwig}.
Additional jets coming from initial and final state radiation (ISR and FSR)  
might fall among the six leading jets and populate the tails in the $m_t^{reco}$  distribution.
These effects are studied using pseudo-experiments where we extract top quark
reconstructed masses from templates generated with different values of $\Lambda_{QCD}$ and scale factor $K$\,\cite{cdfmass}, and comparing them with
the standard templates. Since the shift is very small, we assign the statistical uncertainty on the shift to be the ISR/FSR systematic
 uncertainty.
\par
The choice of parton distribution functions (PDF) inside the proton affects the kinematics of $t \bar t$ events
and thus possibly the top quark mass measurement. We estimate the uncertainty from the difference in top quark mass resulting from the use of Monte Carlo samples 
based on the default CTEQ5L \cite{CTEQ} PDF and on the one calculated from the MRST group \cite{MRST}, MRST72 and MRST75, which 
differ by the value of $\Lambda_{QCD}$ used to compute the PDF.
\par
The background normalization is known to $5\%$ from the tag rate parametrization technique. We vary the background contribution using pseudo-experiments where we
increase or decrease the expected background amount by its uncertainty.
\par
We consider also the uncertainty associated with the small presence of signal in the data-driven background.
To do so we build two background templates where we subtract from the background mass 
distribution the expected  signal mass distribution 
assuming the two values $172.5$ GeV/c$^2$ and $177.5$ GeV/c$^2$ for the top quark mass.
We reconstruct the top quark mass from the two background templates and take the difference in the results as the uncertainty associated to this effect.
A systematic uncertainty due to the finite size of the Monte Carlo samples used to determine the mass templates is determined by varying each of the template bin entries randomly consistent with a Poisson distribution, creating 100 such new templates, reparametrizing them to determine as many fitted top quark masses. The width of the resulting  distribution 
is used as the systematic uncertainty.
\par
A bias in the measurement can arise if an inadequate functional form is used 
 for the $m_t^{reco}$ templates. This is checked 
performing pseudo-experiments where we extract mass values directly from the $m_t^{reco}$ histograms and compare them
with the parametrized p.d.f.'s. The average of the difference between the fitted top quark masses and the input masses is 
chosen as a systematic uncertainty on the functional parametrization.
The $b$-tagging efficiency agrees well between data and simulation; still a possible dependence on jet kinematical properties
could lead to a shift in the measured mass. We evaluate here a systematic due to $E_T$ dependence of $b$-tagging scale 
factors allowing for a slope on the $E_T$ dependence (consistent at  $1 \sigma$ with the measurements), and determining the shift in the fitted top quark mass using the modified templates.
Since the background estimate is data-driven, the analysis is sensitive to an overall
 uncertainty in the $b$-tagging scale factor only 
through signal shapes.
\par
In Table \ref{tab:syst} is shown a summary of all the systematic uncertainties; the total systematic uncertainty amounts to $4.8$ GeV/c$^2$.

\begin{table}[hbtp]
\begin{tabular}{l|c|c}
\hline
\hline
Source & Uncertainty (GeV/c$^2$)\\
\hline
Jet energy scale     &   $4.5$ \\
Generator            &   $1.0$ \\
b-jet energy scale   &   $0.5$ \\
Parton Distribution Function   &   $0.5 $ \\
Background shape     &   $0.5$ \\
Background fraction  &   $0.5$ \\
ISR                  &   $0.5$ \\
FSR                  &   $0.5$ \\
b-tag                &   $0.5$ \\
MC statistics        &   $0.1$ \\
Template parametrization  &   $0.1$ \\
\hline
Total                &   $4.8$ \\
\hline
\hline
\end{tabular}
\centering
\caption[c 2] {Breakdown of systematic uncertainties from different sources.}
\label{tab:syst}
\end{table}

\section{\label{sec:mass}Mass measurement}

After the kinematical selection with $N_{out}\ge 0.91$, the $b$-tagging requirement and the cut on the goodness
of the fit, $\chi^2 \le 16$, we find 926 tags in 772 events.
The background, corrected as in the cross section measurement for the contamination of  $t\bar t$ events (see Section\,\ref{sec:optim}), amounts to $567 \pm 28$ tags.

\begin{figure}
\begin{center}
\includegraphics[width=8.0cm]{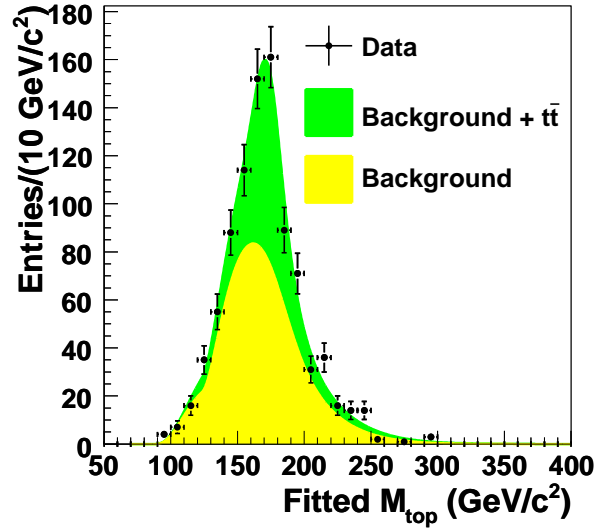}
\caption{The fitted top quark mass distribution for events with
 $N_{out} \ge 0.91$, $\chi^2\le 16$ and at least 1 $b$-tagged jet. 
Superimposed are the background and the $t \bar t$ signal expected for $M_{\rm top} = 174.0$ GeV/c$^2$.}
\label{fig:topmeas}
\end{center}
\end{figure}

\begin{figure}
\begin{center}
\includegraphics[width=8.0cm]{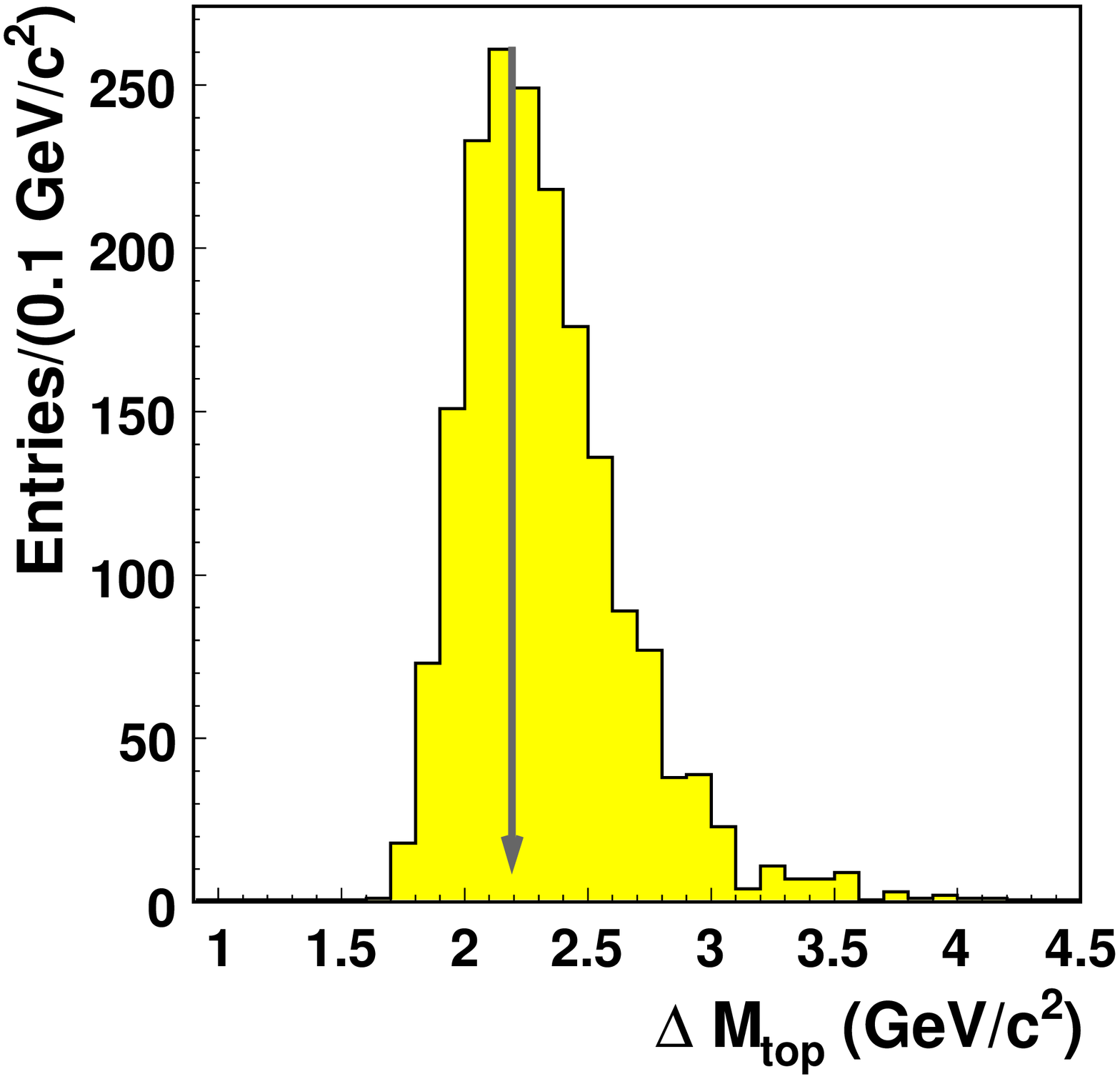}
\caption{The expected statistical uncertainty using
 pseudo-experiments with input top quark mass equal to 174.0 GeV/c$^2$ for the $t\bar t$ signal. 
The arrow represents the measured statistical uncertainty.} 
\label{fig:sigma}
\end{center}
\end{figure}

The likelihood fit is applied to the data sample to derive 
$M_{\rm top} = 174.0 \pm 2.2 (stat.) \pm 4.8 (syst.)$ GeV/c$^2$. 
The plot in Fig.\,\ref{fig:topmeas} shows the fitted top quark mass distribution for the data compared to the expected background and the signal for a top quark mass of 174.0 GeV/c$^2$.
The plot in Fig.\,\ref{fig:sigma} compares the measured statistical uncertainty with the expected distribution from 
pseudo-experiments using as input mass $M_{\rm top} = 174.0$ GeV/c$^2$. We find that the p-value for our statistical uncertainty is
 $40\%$. 
As for the case of the cross section measurement, Monte Carlo simulations indicate the need to increase the statistical uncertainties by 
5\% to account for the use  of multiple entries, one for each tag in the same event, and their correlations.
\par
As a last check, we perform the measurement removing from the likelihood definition
the Gaussian term which constrains the number of background tags to be as predicted via the tag rate parameterization, 
and we obtain nearly the same value, $M_{\rm top} = 174.1 \pm 2.2 (stat.)$ GeV/c$^2$. 
\par
The measurement presented here is the most precise measurement to date of the top quark mass in the all-hadronic decay channel. 
The result is consistent with the measurements obtained in the same channel at $\sqrt{s}=1.8$ TeV 
\cite{tophadPRL},  with the measurement obtained at $\sqrt{s}=1.96$ TeV using ${\cal L}_{int}=311$ pb$^{-1}$
 of data with a different technique
\cite{ideogram}, and with the results obtained in the other channels by CDF\,\cite{cdfmass} 
and D\O\,\cite{d0mass} Collaborations.

\section{\label{sec:fine}Conclusions}

Using an optimized neural-network-based kinematical selection and a $b$-jet identification technique, 
we are able to improve the S/B of the initial multijet sample obtained with a dedicated trigger from 1 in 1100 to approximately 1 in 2. 
With the selected sample, we measure the  $\ttbar$  production cross section  to be 
$\sigma_{\ttbar}=8.3\pm 1.0({\rm stat.})^{+2.0}_{-1.5}({\rm syst.})\pm 0.5({\rm lumi.})$\,pb
assuming $M_{\rm top}=175$\,GeV/$c^2$, improving the previous measurement\footnote{$\sigma_{\ttbar}=7.5\pm 2.1({\rm stat.})^{+3.3}_{-2.2}({\rm syst.})^{+0.5}_{-0.4}({\rm lumi.})$\,pb.} in the same channel\,\cite{tophadPRD}. 
These results agree well with the standard model expectation of $\sigma_{\ttbar}=6.7$\,pb for the 
same value of the top quark mass and with the results obtained in the leptonic channels. 
Using  a slightly modified selection we reconstruct the top quark invariant mass for the overconstrained kinematical system and 
compare it to parameterized templates representing signal and background.
A likelihood fit is used to measure a top quark mass of $M_{\rm top}=174.0\pm 2.2({\rm stat.})\pm 4.8({\rm syst.})$\,
GeV/$c^2$, which improves the previous measurement\footnote{$M_{\rm top}=177.1\pm 4.9({\rm stat.})\pm 4.7({\rm syst.})$\,
GeV/$c^2$.} in the same channel\,\cite{ideogram}, and  agrees well with the results obtained in the leptonic channels. 

%%uncomment for PRD
%%%\begin{acknowledgments}
\section{Acknowledgments}

We thank the Fermilab staff and the technical staffs of the participating institutions for their vital contributions. 
This work was supported by the U.S. Department of Energy and National Science Foundation; the Italian Istituto Nazionale di Fisica Nucleare; the Ministry of Education, Culture, Sports, Science and Technology of Japan; 
the Natural Sciences and Engineering Research Council of Canada; the National Science Council
 of the Republic of China; the Swiss National Science Foundation; the A.P. Sloan 
Foundation; the Bundesministerium f\"ur Bildung und Forschung, Germany; the Korean Science
 and Engineering Foundation and the Korean Research Foundation; the Particle Physics and 
Astronomy Research Council and the Royal Society, UK; the Russian Foundation
 for Basic Research; the Comisi\'on Interministerial de Ciencia y Tecnolog\'{\i}a, Spain; 
in part by the European Community's Human Potential Programme under contract HPRN-CT-2002-00292; and the Academy of Finland. 

%%%\end{acknowledgments}


\begin{thebibliography}{99}
\bibitem{coordinate} We use a cylindrical coordinate system where $\theta$ is the polar angle to the
 proton beam direction at the event vertex, $\phi$ is the azimuthal angle about the beam axis, and
 pseudorapidity is defined $\eta = - \ln \tan(\theta/2)$.
 We define transverse energy as $\Et = E \sin\theta$ and transverse momentum as $p_T = p\sin\theta$ 
where $E$ is the energy measured in the calorimeter and $p$ is the magnitude of the momentum measured by the spectrometer.
\bibitem{met} The missing transverse energy $\vec{\met}$ is calculated as the vector sum of the energy 
in each calorimeter tower multiplied by a unit vector in the azimuthal direction of the tower. If isolated
 high momentum muons are found in the event, the $\vec{\met}$ is corrected by subtracting the
 muon energy in the calorimeter and adding the muon $p_T$ to the vector sum.  $\met$ is defined as 
the magnitude of  $\vec{\met}$.
\end{thebibliography}
\end{document}